\documentclass[namedreferences]{solarphysics}
\usepackage[optionalrh]{spr-sola-addons} 
\usepackage{url}
\usepackage{subfig}
\usepackage{cite}
\usepackage[pdftex]{graphicx}
\usepackage{sidecap}
\usepackage{epstopdf}
\usepackage[pdfborder={0 0 0 },urlcolor=blue,breaklinks]{hyperref}
\usepackage{solaheader}	
\newcommand{\solareceiveddate}{1 August 2012} 
\newcommand{\solaaccepteddate}{22 December 2012}
\renewcommand{\solayear}{2013}

\def\etal{{\it et al.}}

\newcommand{\ion}[2]{#1\,{\sc #2}}
\newcommand{\arcsec}{\hbox{$^{\prime\prime}$}}

\begin{document}
\begin{article}
\begin{opening}
\title{Statistical Analysis of Small Ellerman Bomb Events}
\author{C.J.~\surname{Nelson}$^{1,2}$\sep J.G.~\surname{Doyle}$^{1}$\sep R.~\surname{Erd\'{e}lyi}$^{2}$\sep
Z.~\surname{Huang}$^{1}$\sep M.S.~\surname{Madjarska}$^{1, 3}$\sep M.~ \surname{Mathioudakis}$^{4}$\sep S.J.~\surname{Mumford}$^{2}$\sep K.~\surname{Reardon}$^{4,5,6}$
       }
\runningauthor{C.J. Nelson \etal}
\runningtitle{Small-Scale Ellerman Bomb Events}
   \institute{              $^{1}$ Armagh Observatory, College Hill, Armagh, UK, BT61 9DG \\
				 $^{2}$ Solar Physics and Space Plasma Research Centre, University of Sheffield, Hicks Building, Hounsfield Road, Sheffield, UK, S3 7RH\\
				$^{3}$ UCL-Mullard Space Science Laboratory, Holmbury St Mary, Dorking, Surrey, UK, RH5 6NT\\
				$^{4}$ Astrophysics Research Centre, School of Mathematics and Physics, Queen's University, Belfast, UK, BT7 1NN\\
				$^{5}$ INAF – Osservatorio Astrofisico di Arcetri, I-50125 Firenze, Italy\\
				$^{6}$ National Solar Observatory / Sacramento Peak∗ , P.O. Box 62, Sunspot, NM 88349, U. S. A.\\
                     email: c.j.nelson@sheffield.ac.uk\\ 
             }
\begin{abstract}
The properties of Ellerman bombs (EBs), small-scale brightenings in the H$\alpha$ line wings, have 
proved difficult to establish due to their size being close to the spatial resolution of even the 
most advanced telescopes. Here, we aim to infer the size and lifetime of EBs using high-resolution data 
of an emerging active region collected using the { \it Interferometric BIdimensional Spectrometer} (IBIS) 
and { \it Rapid Oscillations of the Solar Atmosphere} (ROSA) instruments as well as the { \it Helioseismic and Magnetic Imager} (HMI) onboard the { \it Solar Dynamics Observatory} (SDO). We develop an algorithm to track EBs through their evolution, finding that EBs can often be much 
smaller (around $0.3$\arcsec) and shorter lived (less than $1$ minute) than previous 
estimates. A correlation between G-band magnetic bright points and EBs is also found. Combining SDO/HMI and G-band data gives a good proxy of the polarity for the vertical magnetic field. It is found that EBs often occur both over regions of opposite polarity flux { and strong unipolar fields}, possibly hinting at magnetic reconnection as a driver of these events.The energetics of EB events is found to follow a power-law distribution in the range of ``nano-flare'' ($10^{22-25}$ ergs). 
\end{abstract}
\end{opening}

\section{Introduction}
     \label{S-Introduction}
Emerging active regions are some of the most diverse and abundant regions to study within the photosphere. Active regions possess a wide range of both large-scale ({\it e.g.} sunspots and pores) and small-scale ({\it e.g.} Ellerman bombs, also known as EBs) structures which, in a way that is yet to be fully understood, interact to form the complex overlying chromosphere. Here we study EBs, small-scale brightenings in the wings of the Balmer H$\alpha$ line profile, which occur in regions of high magnetic activity especially near opposite-polarity regions (see {\it e.g.} \opencite{Georgoulis02}, \opencite{Pariat04}). 

Small-scale structures in the solar atmosphere have proved difficult to study due to their size being 
close to the spatial resolution of current observational instruments. EBs, possible 
magnetic reconnection events in the { lower atmosphere} due to emerging flux, are one such example. First reported 
by \inlinecite{Ellerman17} and originally named ``solar hydrogen bombs'', EBs have been reported to have a size 
of the order of one arcsecond (see {\it e.g.} \opencite{Kurokawa82}) and lifetime of around $10-15$ 
minutes on average (see {\it e.g.} \opencite{Vorpahl72}, \opencite{Roy73} and more recently 
\opencite{Watanabe11}). The energetics of EBs have been estimated by \inlinecite{Georgoulis02} and \inlinecite{Fang06}, finding that each EB has a total lifetime energy of around $10^{27}$ ergs, a value typically associated with ``micro-flares''. It has been widely acknowledged ({\it e.g.} \opencite{Georgoulis02}) that these 
estimates are upper bounds which will be revised occasionally by higher resolution, higher cadence 
data such as we present here. 

The majority of investigations of EBs have { been} taken place the H$\alpha$ line wings. EB events, however, are also seen in other wavelengths such as the $1600$ \AA\ continuum and the \ion{Ca}{ii} wings (see, for example, \opencite{Herlender11}). Recently, \inlinecite{Qiu00} suggested that over $50\ \%$ of EBs show a correlation to UV 
emissions in the $1600$ \AA\ continuum; a similar result was also found by \inlinecite{Pariat07}, providing evidence that EBs are upper-photospheric, lower-chromospheric events. 

Recently, a link has been suggested between EBs and G-band magnetic bright points (MBPs) (for a discussion of MBPs see \opencite{Jess10b}), which are ubiquitous in the solar photosphere. MBPs are the locations where the magnetic flux clumps
together to form small magnetic concentrations with field strengths of the order of a kiloGauss (\opencite{Stenflo85}). Their increased brightness
is due to the reduced pressure and opacity within the flux tube allowing the observer to view a deeper, hotter region of the photosphere as well as
heating of the plasma within the flux tube by material surrounding its walls. The increased temperature reduces the abundance of the CH molecule,
thus making MBPs appear brighter in G-band imaging (\opencite{Shelyag04}).  \inlinecite{Jess10} has recently suggested that the interaction of
neighbouring MBPs can lead to multiple EBs as a consequence of forced reconnection{,} thus hinting at a link between EBs and MBPs.

\begin{figure}
\centering
\subfloat{\includegraphics[scale=0.25]{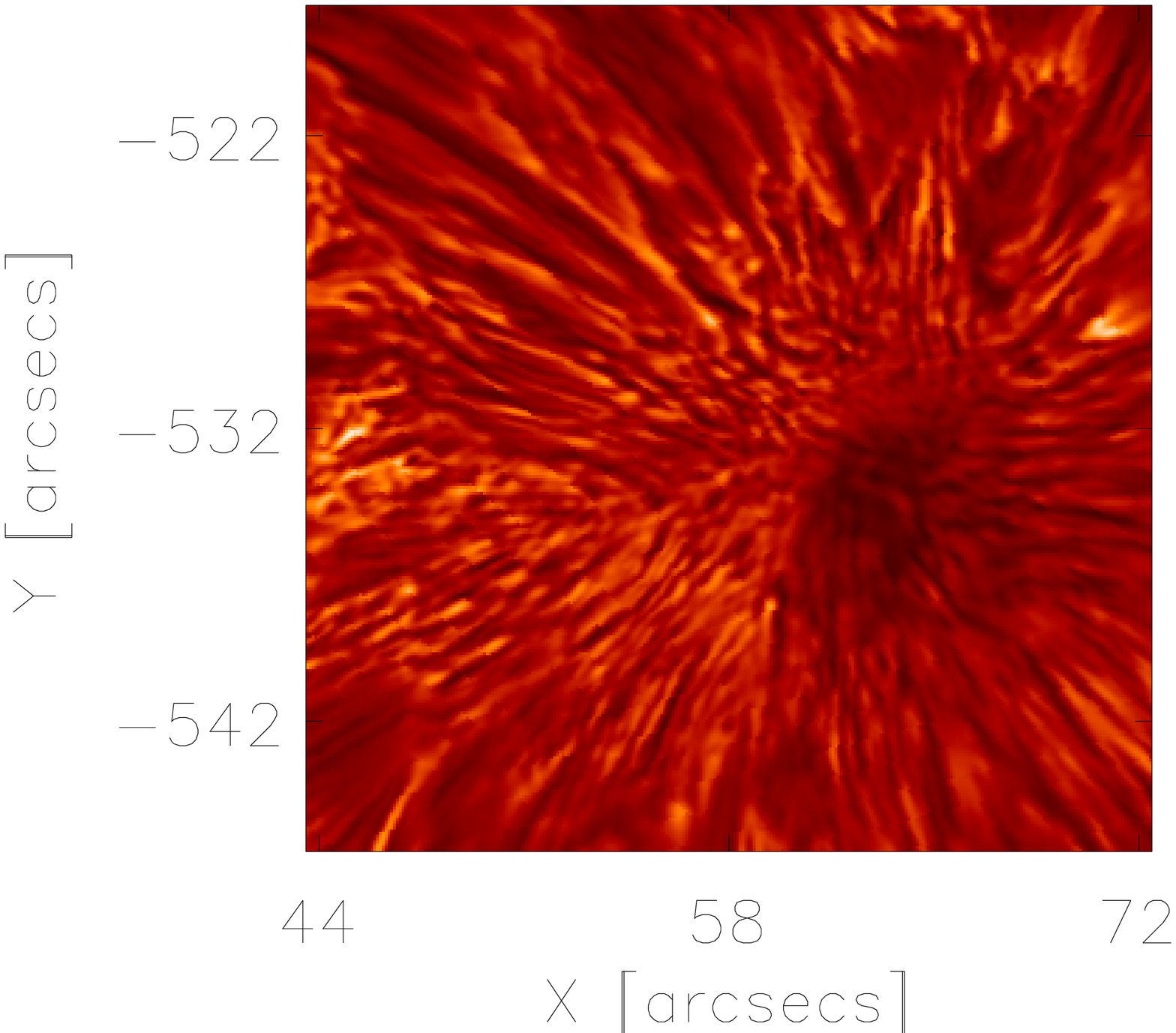}}
\hspace{-36pt}
\subfloat{\includegraphics[scale=0.25]{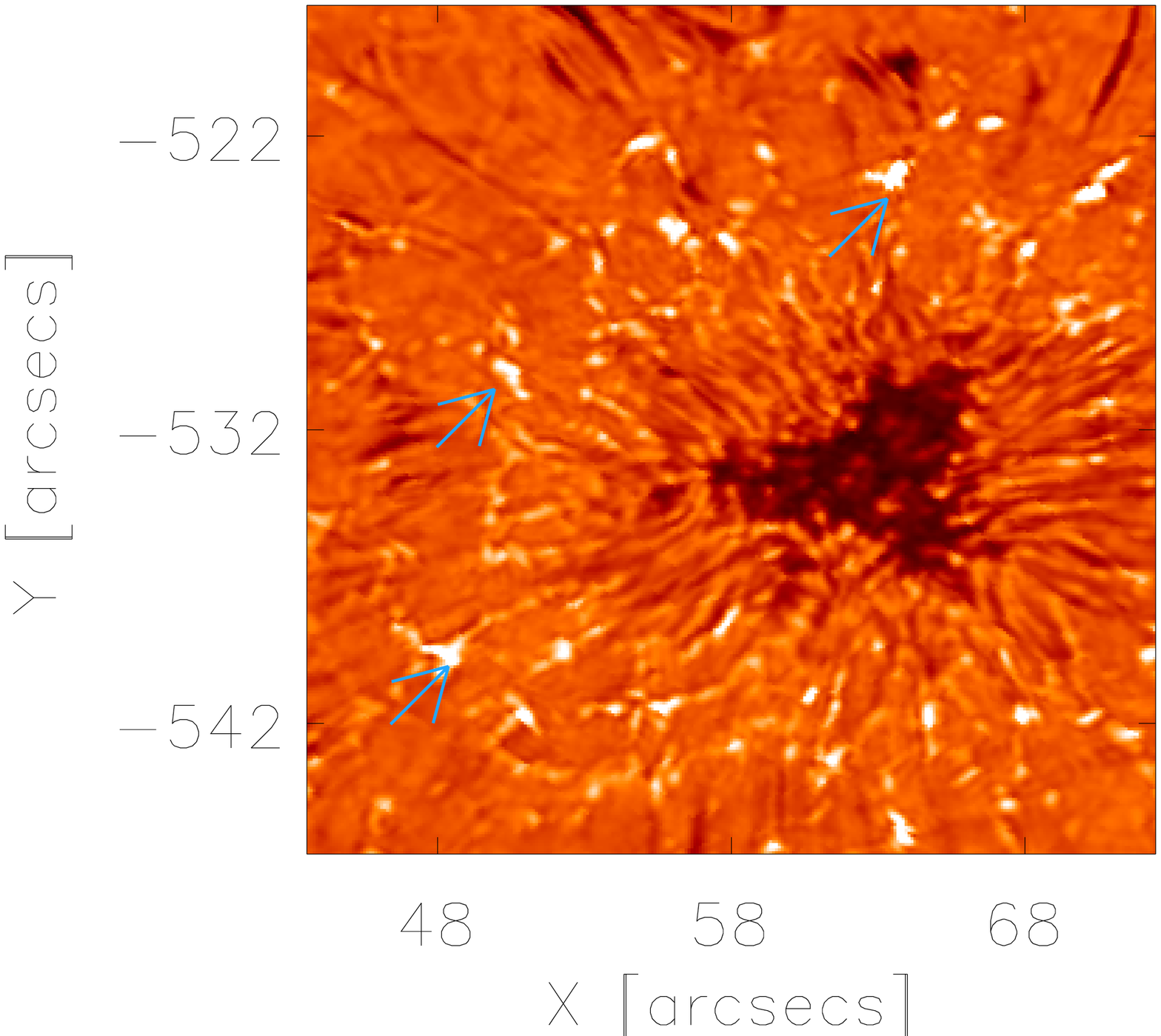}}
\vspace{-10pt}

\subfloat{\includegraphics[scale=0.25]{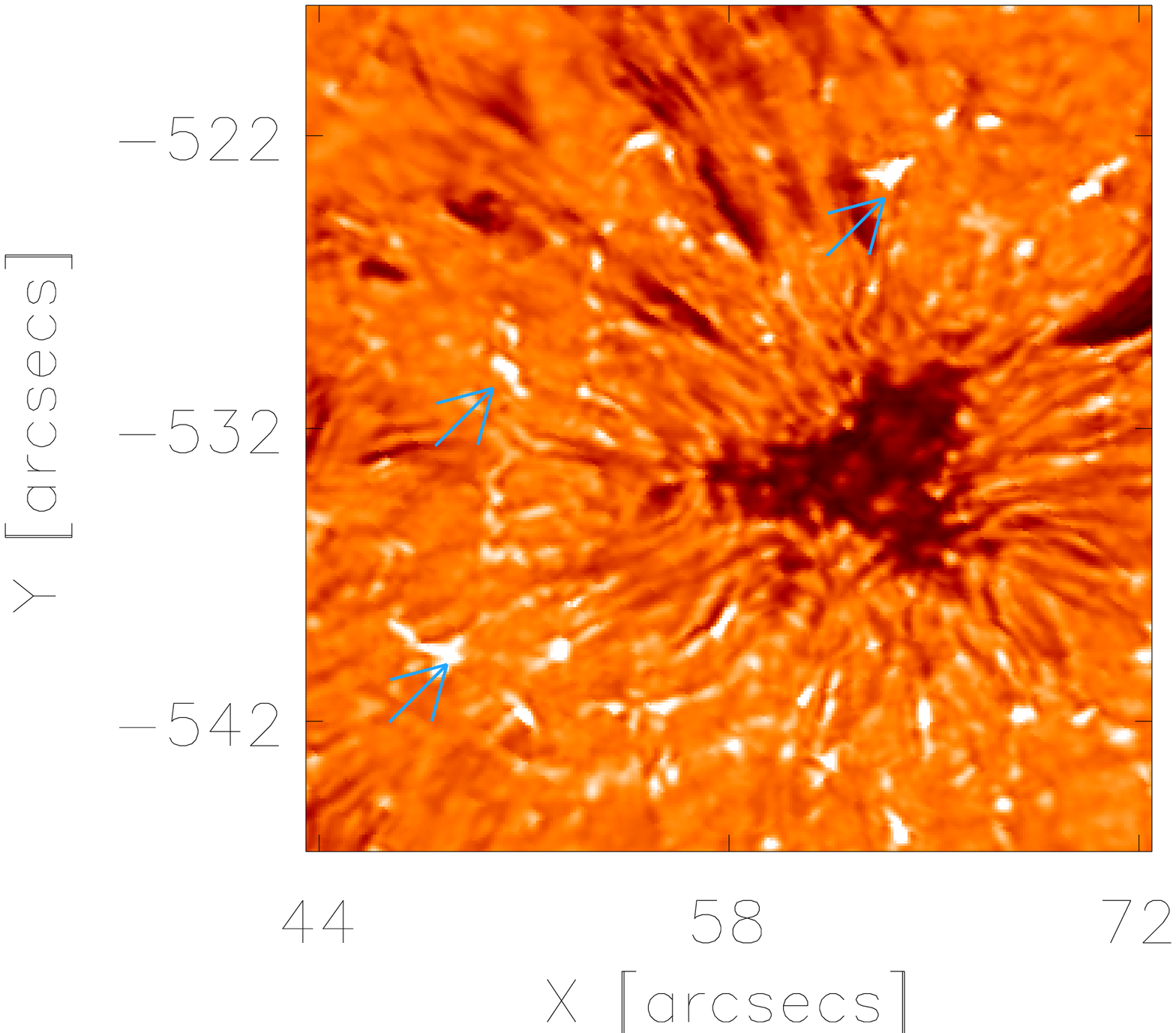}}
\hspace{-36pt}
\subfloat{\includegraphics[scale=0.25]{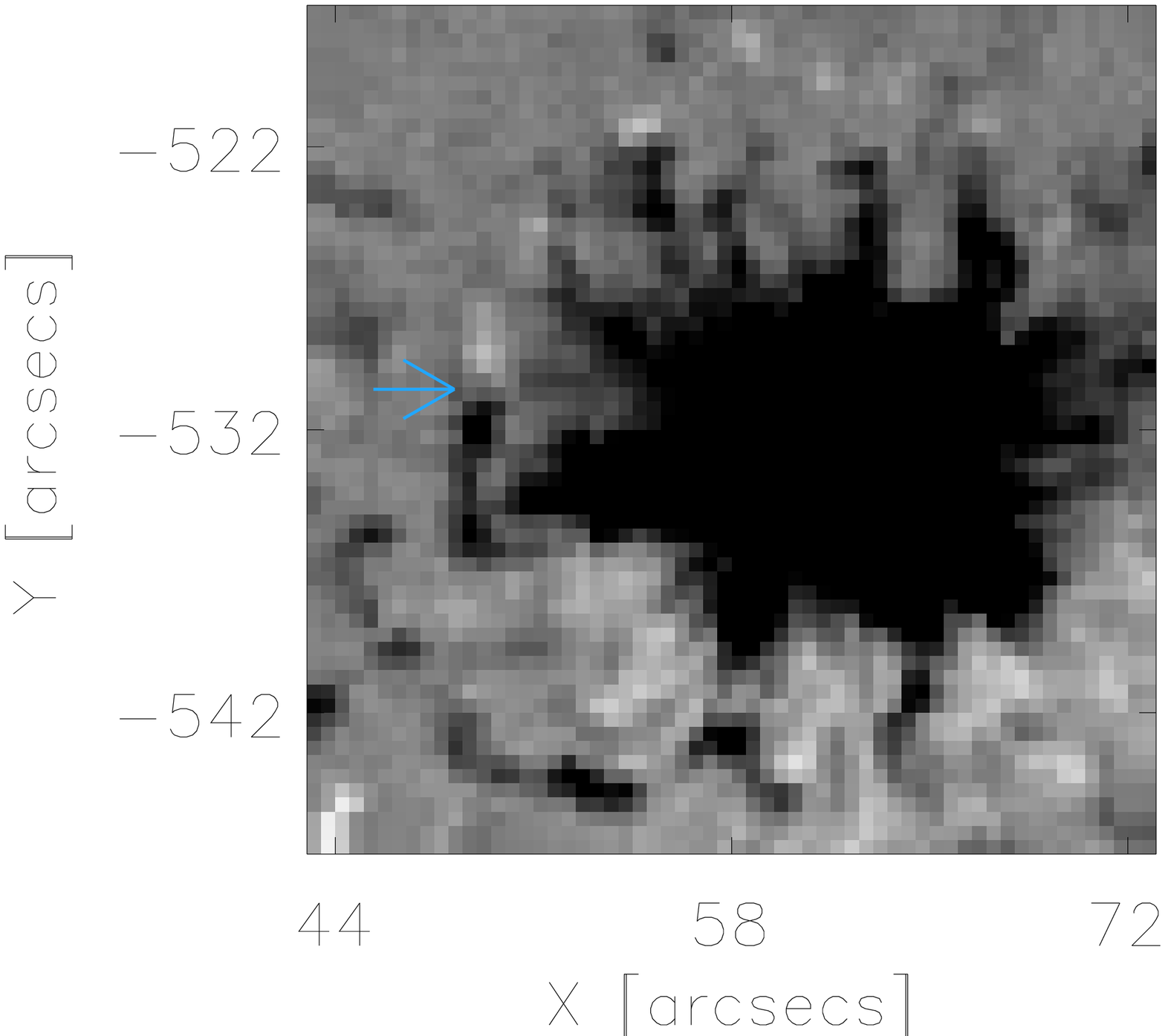}}
\caption{Images of Active Region NOAA 11126 (taken at 15:02:49) (a) at the H$\alpha$ line centre as well as (b)-(c) at $\pm0.7$ \AA, observed with DST/IBIS. Three EBs are shown with arrows. The same region is shown in (d), detected by the SDO/HMI instrument (using a threshold of $\pm{400}$ Gauss), which has been spatially and temporally aligned with (a-c).}
\end{figure}

There have been a number of recent studies both on observations (\opencite{Georgoulis02}, \opencite{Pariat07}, \opencite{Watanabe11}) and modelling (\opencite{Isobe07}, \opencite{Archontis09}) of  
EBs. The main focus of many of these studies has been to provide evidence for magnetic 
reconnection as the possible formation mechanism of EBs. A good cartoon was presented by 
\inlinecite{Georgoulis02}, which depicts the possible magnetic topologies that could lead to magnetic 
reconnection in the { lower-atmosphere} either through emerging flux in bipolar regions or topologically complex unipolar fields. In this work, we find evidence that EBs are formed over regions of strong magnetic field, shown in both SDO/HMI data and the G-band continuum.

Magnetic reconnection is largely associated with high-energy events in the corona such as coronal mass ejections (CMEs) and large flares; however, in the past two decades, the applicability of the idea of magnetic reconnection to the { lower-atmosphere} has also been discussed. The process was described in \inlinecite{Litvinenko99}, who suggested that reconnection was likely to occur mainly around the temperature minimum region at $z\approx{600}$ km; this value does not appreciably disagree with the initial height of $z\approx{400}$ km found by \inlinecite{Harvey63}. In \inlinecite{Litvinenko07} it was suggested that { lower-atmospheric} magnetic reconnection is an important precursor that could lead to larger events in the corona as well as mass supply into the chromosphere and corona through the ejection of cool photospheric plasma in the form of filaments.

EBs were originally thought to be precursors for flares; however, \inlinecite{McMath60}
found this not to be the case through an analysis of a large array of data collected using a multitude 
of instruments in the preceding years. They did find, however, that EBs are mainly observed around 
the penumbra of active region sunspots; this has been discussed more recently by \inlinecite{Pariat04} who found that the majority of EBs occur in the trailing plage region. In this article we shall discuss the 
positioning of EBs identified in our dataset, finding formation to be common in both the unipolar penumbra and bipolar regions, in the surrounding active region close to the spot, suggesting both regions may have the required background conditions to form EBs. \inlinecite{Madjarska09} studied an H$\alpha$ surge, finding that the event was triggered by EBs forming in a similar spatial position. This is important as it could possibly suggest a coupling between the photospheric EB events and chromospheric H$\alpha$ surges.

We structure our study as follows: in Section 2 we discuss our observations and data reduction methods; Section 3 presents our results along with the definition we have used to calculate our statistics. Section 4  contains our conclusions and poses some questions that should be answered through further study.

\section{Observations}
     \label{S-Observations}
In this article, we make use of high-resolution data collected by the { \it Interferometric BIdimensional Spectrometer} (IBIS) and { \it Rapid Oscillations of the Solar Atmosphere} (ROSA) instruments at the { \it Dunn Solar Telescope} (DST), 
{ \it National Solar Observatory, Sacramento Peak, NM}, between 15:02 UT and 17:04 UT on 18 November 2010. We limit our analysis 
to the first $90$ minutes of observations due to the deterioration of the seeing in the subsequent images ({\it i.e.} 
leaving $200$ IBIS frames). IBIS data have a pixel 
size of $0.096$\arcsec\ and, due to the excellent seeing, an approximate spatial resolution of $0.192$\arcsec. IBIS was set to produce a 
$15$-point { H$\alpha$} line scan (between $\pm{1.4}$ \AA\ from the line centre in unequal steps) and $50$ images in the line centre and both line wings at $\pm0.7$~\AA\ before repeating this 
routine ({\it i.e.} a total of $165$ images per cycle), sampling a total FOV of $96$\arcsec $\times 96$\arcsec. G-band images from the ROSA instrument were collected during the 
same period with a cadence of $0.64$ seconds (by combining $32$ images, each with an intergration time of $0.02$ seconds), a pixel size of $0.059$\arcsec\ and a spatial resolution of $0.118$\arcsec, covering a FOV of $58$\arcsec $\times 58$\arcsec\ (situated entirely within the IBIS FOV). We also use the { \it Helioseismic and Magnetic Imager} (HMI) onboard the { \it Solar Dynamics Observatory} (SDO) spacecraft, which has a pixel size of $0.5$\arcsec\ a spatial resolution 
of $1$\arcsec\ and a cadence of $45$ seconds (whilst observing the whole disc). In Figure 1, we show context images with a field-of-view of 
$29$\arcsec $\times{29}$\arcsec\ (to illustrate the complex region surrounding the sunspot) 
of the emerging active region NOAA $11126$ at $31$S latitude $00$ longitude with co-aligned HMI and IBIS images. The DST tracked the leading spot, which evolved slowly through the time series. This spot was linked, by a coronal-loop arcade, to two small opposite-polarity spots situated in the trailing plage region.

Data analysis was conducted after two key steps: the speckle-imaging process and the alignment of the datasets. 
The speckle-imaging technique (\opencite{Woger08}) was used for the IBIS line centre and wing data with each repetition 
of the IBIS routine contributing one final image with a cadence of $26.9$ seconds. This process was applied to 
counteract the changes in seeing over the course of a single cycle by combining $50$ short-burst images into a lower-cadence, higher-resolution image. For the ROSA G-band data, 32 images were used giving a final cadence of 0.64 seconds. 
Each H$\alpha$ speckle image was aligned to the previous image to eliminate jitter. Next, alignment of the datasets from separate instruments was carried out by co-aligning three specific reference points 
on each image and then rotating the images accordingly. { Contour plots of the region as observed by DST/IBIS were then overlaid 
on SDO/HMI and DST/ROSA images with large features, such as the sunspot, used to validate the alignment.}

Our H$\alpha$ observations have two main limitations that should be considered before future studies in this area are undertaken. Firstly, there is a cadence of $26.9$ seconds which is large in comparison to the overall lifetime of network bright points (NBPs) as predicted by \inlinecite{Watanabe11} who note that the function $y\propto\exp(-x/C)$ fits the lifetimes and occurrence rates of NBPs well (where $y$ is the number of NBPs for each lifetime, $x$). Therefore, our analysis is limited to events of lifetime longer than $54$ seconds (or two consecutive frames) and could miss events that form and disappear within this time limit (brightenings that occur in one time frame are included in the histogram of lifetimes but are not included elsewhere). In Section 3.1, we discuss these shorter-lived events and how they link to EBs. Secondly, the spatial resolution of our data is potentially too large, at $0.192$\arcsec, to capture all EBs. EBs may form in much smaller areas than this. We do acknowledge that due to the speckling of the IBIS images, short-lived EBs may be missed; however, with the current observational techniques and the small scales of the events studied in this research, the compromise between high cadence and high spatial resolution is important.

\section{Statistical Analysis}
     \label{S-Statistical Analysis}
\subsection{Identification of Ellerman Bombs}
The definition of an Ellerman bomb (and its relationship to network bright points) varies widely 
within the literature on the topic. Here, we shall discuss, and justify, what we have found 
appropriate and used as the definition of an Ellerman bomb event and how this differs from those 
used by other authors. We suggest that, in essence, an Ellerman bomb is a brightening in the line 
wings of the H$\alpha$ optical line (an example is shown in Figure 2), which is also visible in 
the wings of other chromospheric lines such as \ion{Ca}{ii} as well as the $
1600$ \AA\ continuum (\opencite{Qiu00}). In his original article studying the H$\alpha$ line profile, \inlinecite{Ellerman17}, suggested that the brightenings were ``a very brilliant and very narrow band extending four or five \AA\ on either side of the line, but not crossing it.'' This definition still holds, 
suggesting that EBs are features of the upper photosphere and lower chromosphere that are 
covered by the highly opaque network of fibrils and mottles that form the complex chromosphere, which it is possible to observe using the H$\alpha$ line 
centre. 

\begin{figure}
\centering
\subfloat{\includegraphics[scale=0.34]{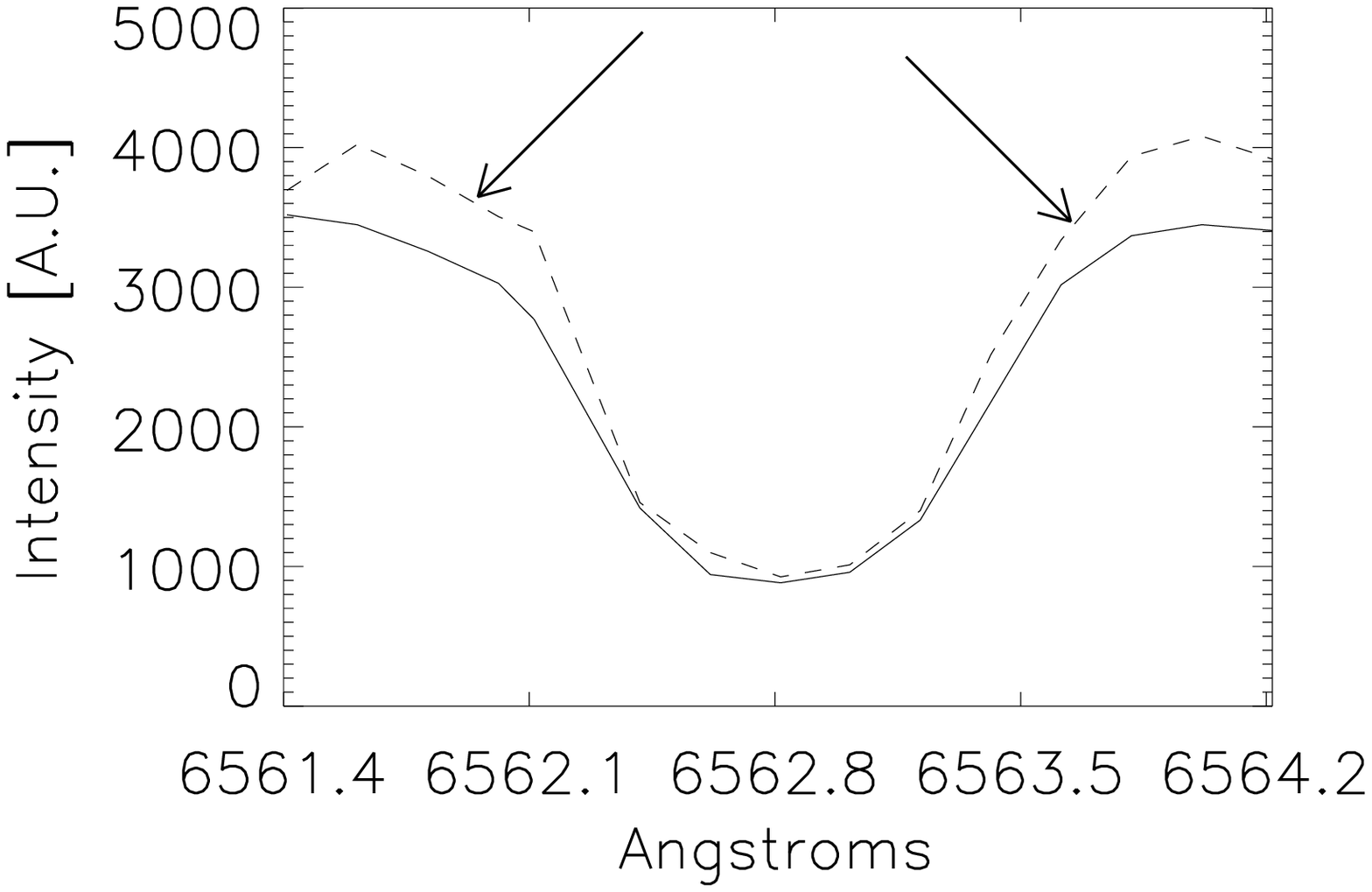}}
\hspace{10pt}
\subfloat{\includegraphics[scale=0.2]{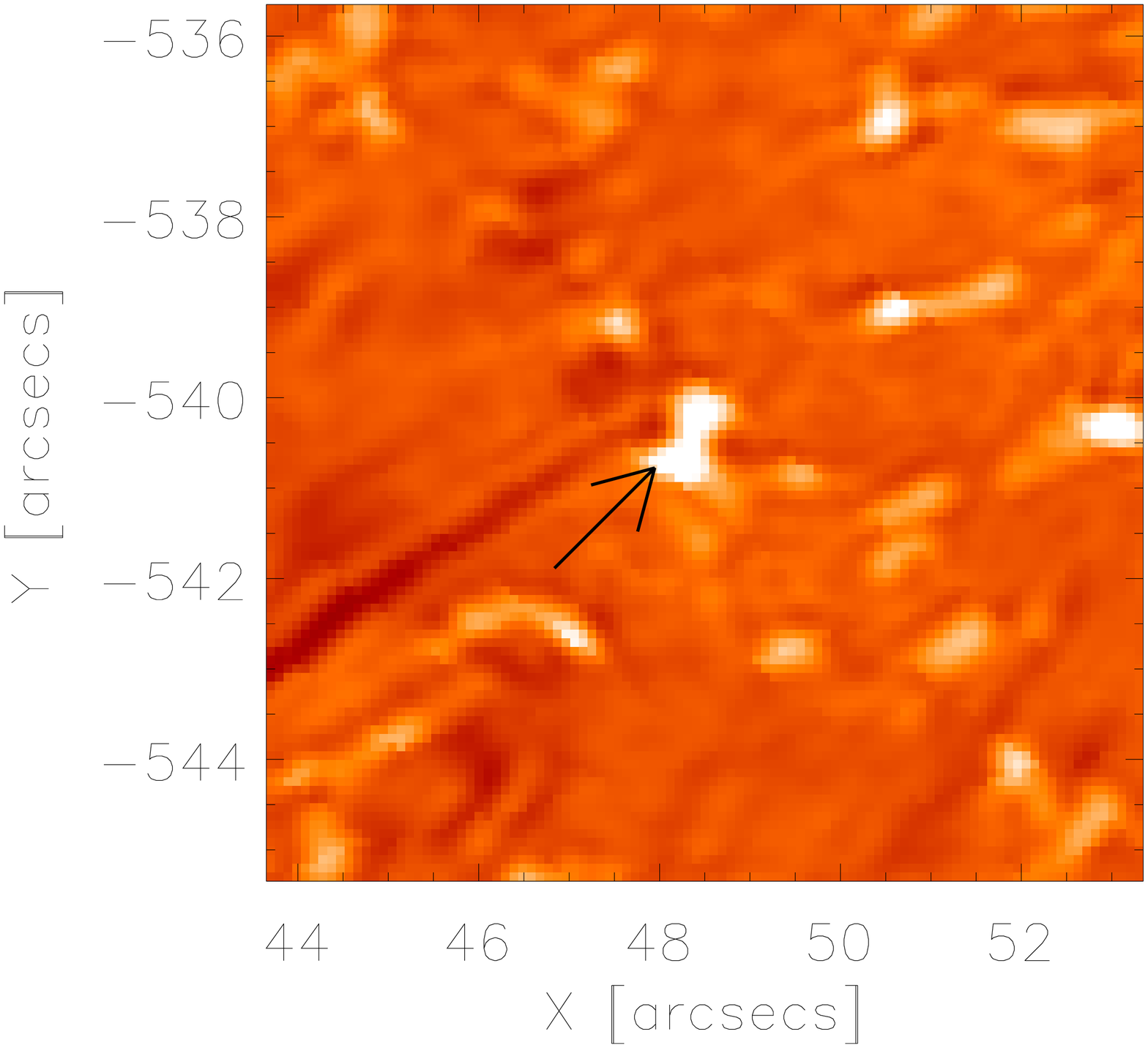}}
\caption{(a) H$\alpha$ line-profile showing an Ellerman bomb stretching out into the line wings.  The solid line is from a quiet region and the dashed line is the profile for the EB indicated by the arrow in the blue wing { image} shown at 15:14:01 by (b). This EB is not taken from the same frame as Figure 1.}
\end{figure}

The threshold at which an Ellerman bomb is no longer classified as a network bright point is also a point of interest. Recently, \inlinecite{Watanabe11} suggested that the difference between EBs and network bright points was both their brightness (EBs are more intense events) and their tendency to flare (both in intensity and size). This definition may, however, cause confusion, as smaller events that possess brightenings may not be observed flaring due to the ``flare'' being within the spatial resolution of the data. Here, we define an Ellerman bomb only in terms of brightening and set our threshold at $130\ \%$ of the average intensity of the image (this value is the upper limit of thresholds discussed in \opencite{Georgoulis02}). This value was chosen as it incorporated all major EB events selected by eye within this dataset as well as limiting the effect of the sunspot on the average FOV intensity.

Overall, we suggest and use the following as parameters for an automated method for the detection of EBs that can be implemented to follow their evolution through time:
\begin{itemize}
\item{The region is over $130\ \%$ of the FOV average brightness in both the red and blue wings of the H$\alpha$ line-profile.}
\item{It has an initial spatial overlap in the red and blue wings.}
\item{Its area is greater than or equal to two pixels.}
\item{If there is a spatial overlap in both the red and blue wings between frames then the EB is deemed to have lived into the next frame. After the initial frame, no overlap is required between the wings as small-scale events may separate within their lifetime.}
\item{If there are multiple events with spatial overlap the one with the largest spatial correlation is taken; the others are neglected.}
\item{Any EBs that occur in the first or last frame are removed from the sample so that only EBs with a complete lifetime are included.}
\end{itemize}
Note that one of the main differences in the definition above from that by \inlinecite{Watanabe11} is that we do not set a lower time limit for the lifetimes of EBs ($240$ seconds is used in by \opencite{Watanabe11}). We discuss this change in Section $3.3$. Any analysis on EB events should also take into account the original description by \inlinecite{Ellerman17}, in that EBs are not visible in the line centre. In this dataset there are no  brightening events observed in the H$\alpha$ line centre during our time series, this was not included in the algorithm; however, this would potentially need to be revised for future studies.

EBs have been predicted to occur at a rate of $1.5$ per minute in an 
$18$\arcsec $\times{24}$\arcsec\ region by \inlinecite{Zachariadis87}. Interpolating 
this to our FOV, we would expect around $30$ EBs to occur per minute. Therefore in our 90-minute dataset  
we should expect to see $2690$ EBs. \inlinecite{Zachariadis87} suggest that EBs can form, fade and recur in the same position meaning that $2690$ could be extremely conservative. By classifying an event that rises above the $130\ \%$ threshold, fades and then rises above the threshold again as two separate events, we present $3570$ EBs in this time range ($133\ \%$ of the predicted value presented by \opencite{Zachariadis87}).

\vspace{2pt}
\begin{figure}
\centering
\subfloat{\includegraphics[scale=0.5]{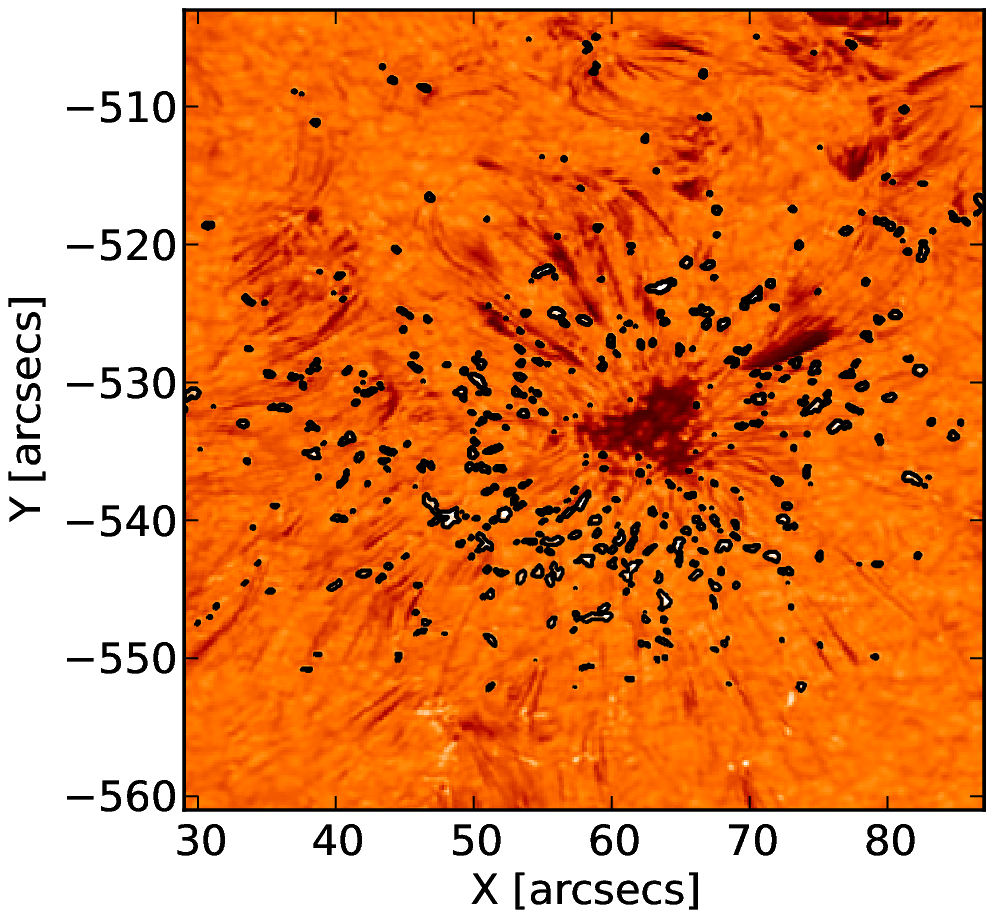}}
\hspace{-20pt}
\subfloat{\includegraphics[scale=0.5]{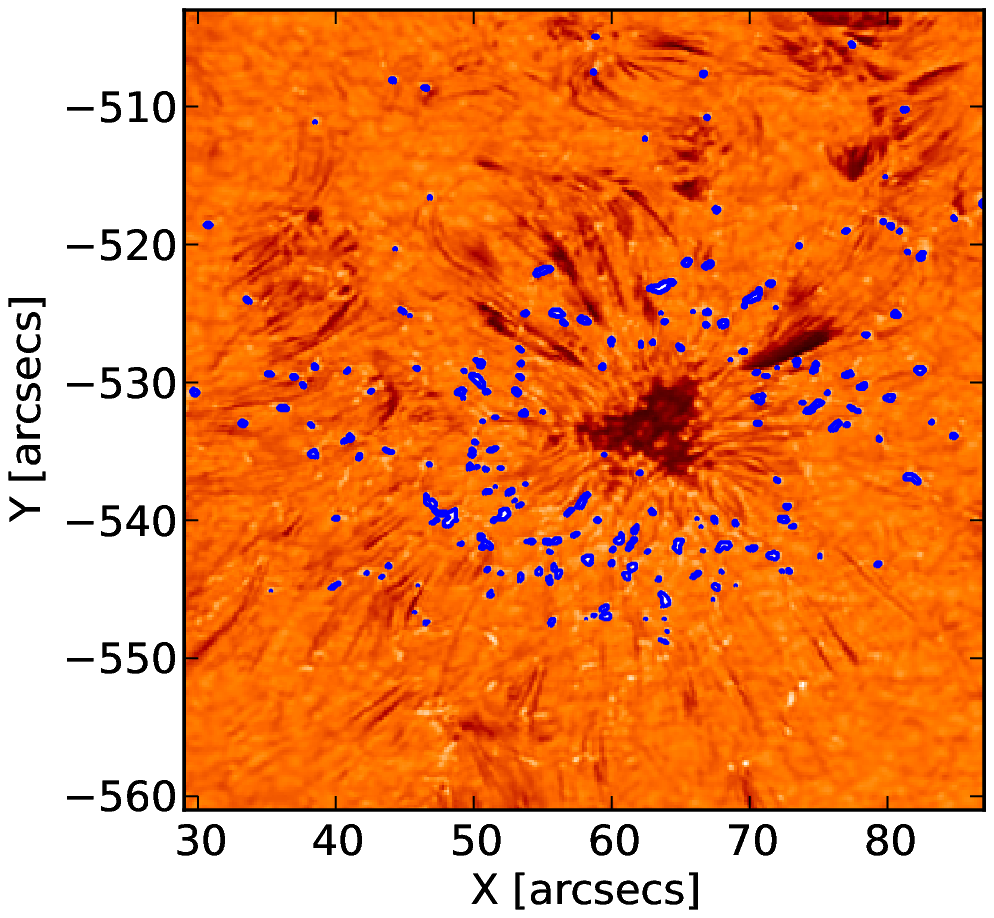}}

\subfloat{\includegraphics[scale=0.5]{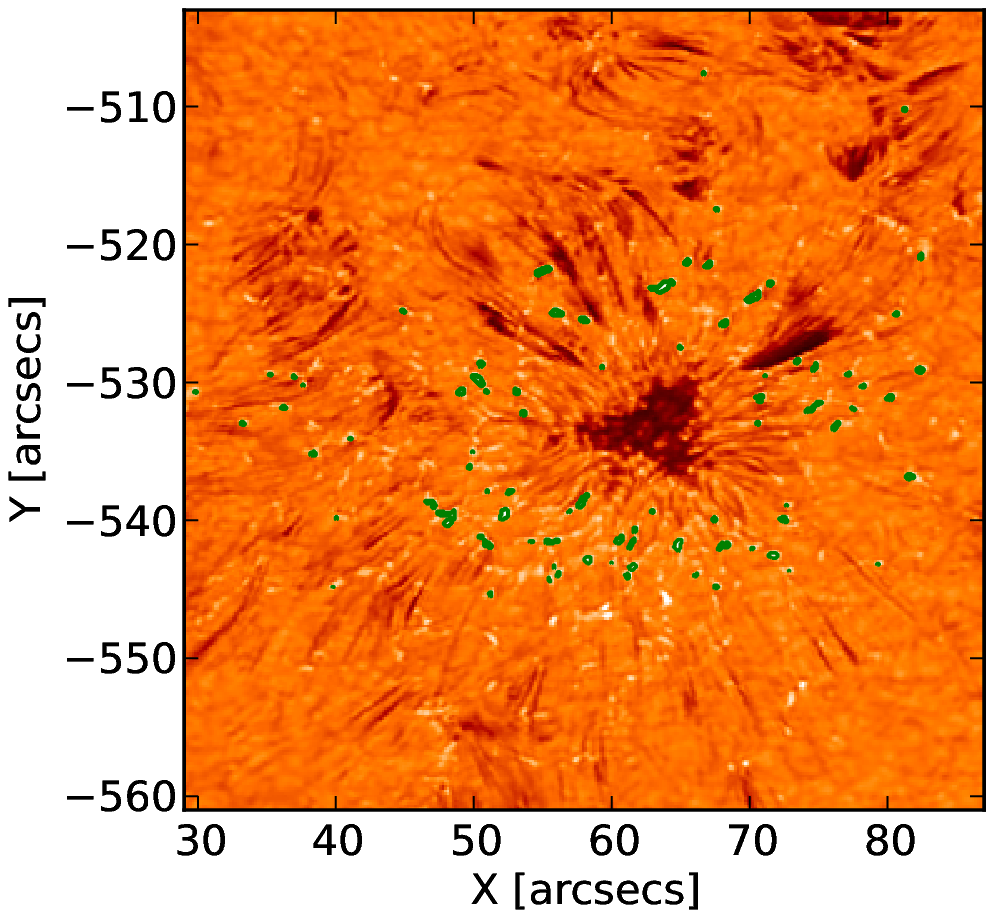}}
\hspace{-20pt}
\subfloat{\includegraphics[scale=0.5]{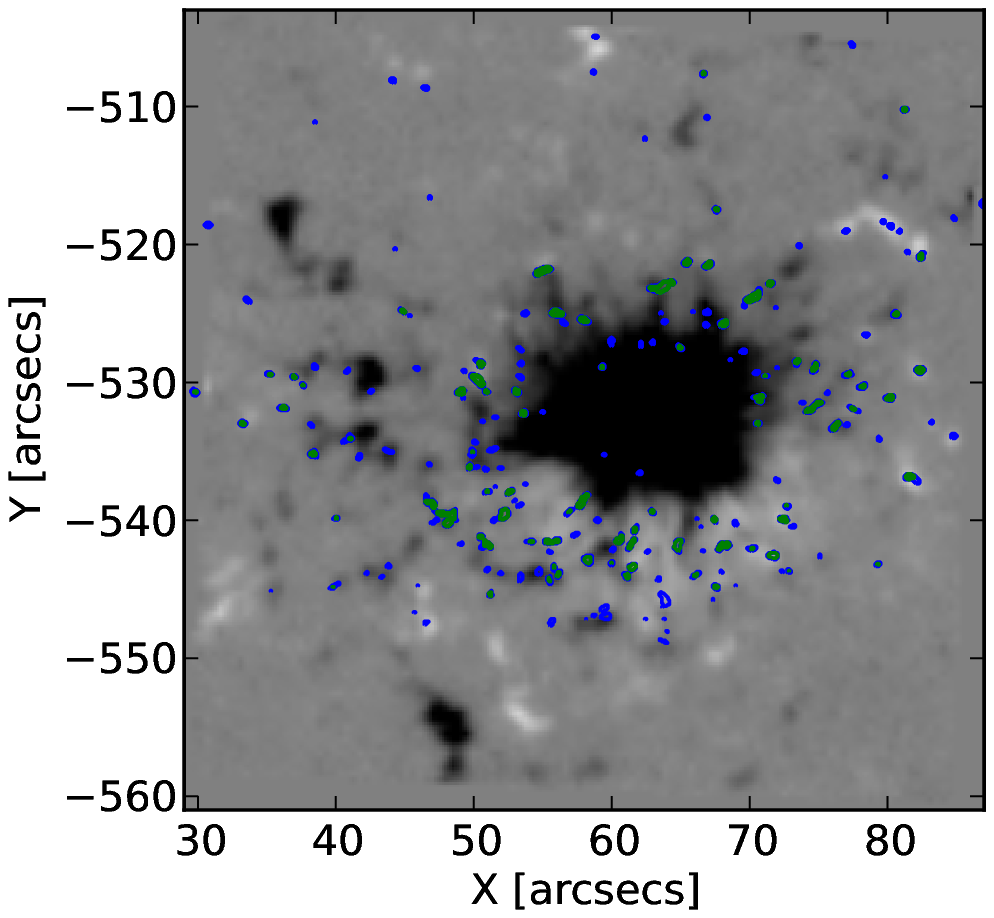}}
\caption{IBIS red wing images with contours of several brightness threshold values overplotted with: (a) $120\ \%$, (b) $130\ \%$ and (c) $140\ \%$, respectively. A temporally aligned HMI image ($\pm{400}$~G) is included in (d) with $130\ \%$ and $140\ \%$ contours overplotted.}
\end{figure}

\subsection{Spatial Occurrence of Ellerman Bombs}
EBs form in highly magnetic regions, for example, in bipolar regions within emerging active regions or over complex unipolar topologies such as the penumbrae of sunspots (\opencite{Georgoulis02}). Figure 3 shows contours of 
the EBs from the same image at different brightness threshold values as well as an HMI image of 
the same region. The temporal alignment between the H$\alpha$ and HMI data is less than $10$ seconds and we 
estimate that the spatial alignment is better than an arcsecond therefore within the spatial resolution of the 
HMI instrument.

The occurrence of EBs in emerging active regions such as NOAA 11126, which we study, has been used as evidence that they are magnetic reconnection events in the { lower-atmosphere} due to emerging flux ({\it e.g.} \opencite{Georgoulis02}). They have, previously, been predominantly observed between the leading spot and the following plage region; however, they have also been reported as occurring { within complex penumbrae, where complex unipolar magnetic fields dominate} (for flux emergence see \opencite{Pariat04}; for a good review of photospheric magnetic reconnection see \opencite{Litvinenko99}). Both of these regions show strong and complex magnetic fields in photospheric magnetograms, therefore implying that magnetic fields are, in some way, responsible for the formation of EBs.

The FOV that we are observing is extremely complex in terms of magnetic field structuring. { For example, we see from Figure 1(d) that there is a region of both positive and negative magnetic field indicated by the arrow to the left ($51$\arcsec, $-530$\arcsec) of the spot, which leads an intense brightening indicated by an arrow in Figures 1(b) and (c).  Also, around the spot sits a large, penumbral structure as well as both small and large fibril structures (easily seen in Figure 1(c)).}

We notice in Figure 3 that brighter EBs tend to form nearer to the strong magnetic fields { within} the penumbra of the leading spot. With a threshold of $120\ \%$ we find a near-ubiquitous covering of EBs in the FOV, especially trailing the spot (to the left). { As has been discussed previously, many of these events appear to be small, localised changes in the background intensity, which are selected by the algorithm due to the decreasing of the average FOV intensity caused by the sunspot. By increasing the internsity threshold to $130\ \%$, many of the contours created at $120\ \%$ disappear and leave what, by eye, appears to be all major EB events. A threshold of $140\ \%$ focuses the brightenings around the spot at the regions of highest magnetism (Figure 3(c)); however, several examples of large, flaring brightenings are not selected with this threshold, implying that it may be too high.}

The spatial correlation between EBs and strong vertical magnetic field, {  as shown in Figure 3(d)}, is another hint that EBs are in fact magnetic events whose intensity is dominated by the strength of the magnetic field. { By contouring EBs identified with thresholds of $130\ \%$ and $140\ \%$ on a temporally aligned SDO/HMI magnetogram, it is easy to see the correlation between H$\alpha$ brightenings and strong magnetic fields. At approximately ($55$\arcsec, $-543$\arcsec), one can pick out a line of negative polarity magnetic field that is mirrored exactly by EBs. An example at a region of positive polarity can be seen at ($80$\arcsec, $-519$\arcsec). The cadence and, more importantly, the spatial resolution of the HMI images are, however, too low to pick out fine-scale structuring within the magnetic field meaning nothing more than a correlation between EBs and strong, vertical photospheric magnetic fields is supported by this analysis. Higher-resolution magnetograms are required to fully analyse whether EBs are} { associated with separatrix of magnetic structures as suggested by Georgoulis \etal\ (2002).}

\subsection{Lifetime of Ellerman Bombs}
Previous studies of EBs have been limited by relatively long-cadence observations (the 
Flare Genensis Experiment used by \opencite{Georgoulis02}, and \opencite{Pariat04}, for example, 
had a cadence of $\approx {3.5}$ minutes) except for a small number of key studies ({\it e.g.} \opencite{Watanabe11}), which could miss the formation of EBs. Here, we present an analysis of EB lifetimes 
using high-resolution, relatively high-cadence observations. The speckle reconstruction technique 
we applied has improved the spatial resolution of these data, meaning that even brightenings of around { $0.14$\arcsec\ are observable} { with a temporal cadence of $26.9$ seconds}. The resolution of our data in both spatial and temporal coordinates 
allows us to detect EBs both occurring and, more importantly, recurring (see {\it e.g.} \opencite{Georgoulis02}), which influences previous lifetime estimates.

In Figure 4(a), we show a histogram of the lifetimes of EBs for our dataset. We find that the { function $y\propto{\exp(-x/C)}$ (where $y$ and $x$ are the total number and lifetime of EBs respectively and $C=2.7\pm0.16$)} proposed by \inlinecite{Watanabe11} fits our histogram for lifetimes less than $20$ minutes, implying the possibility that EBs could be formed and disappear { on timescales shorter than the temporal resolution presented here}. Even higher cadence is required to fully answer this question. { Although the fit of this curve appears to have a long tail, it should be noted that only seven events do not match the the exponential function, meaning that $5402$ events do fit this curve.}

\begin{figure}
\centering
\subfloat{\includegraphics[scale=0.5]{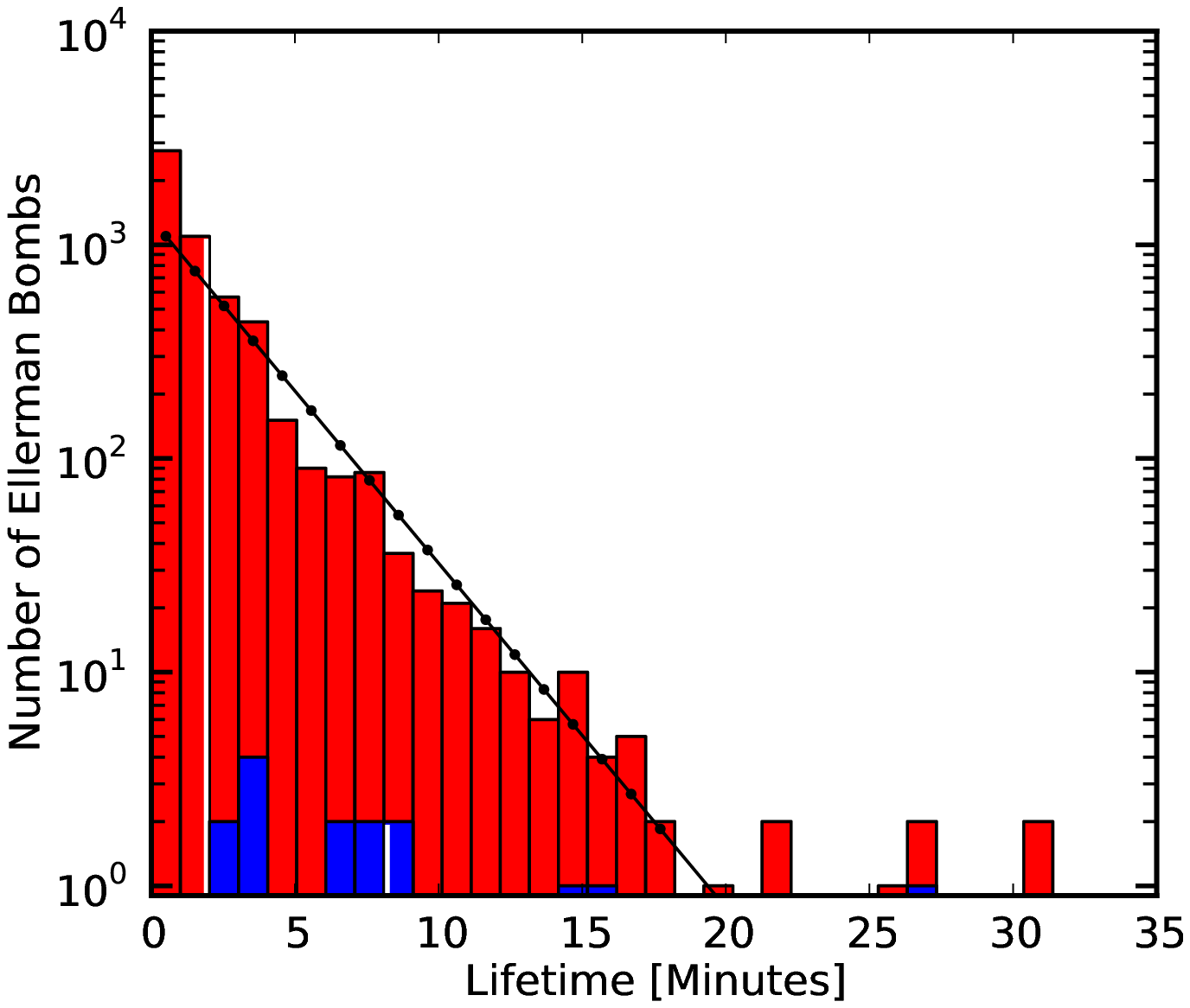}}
\vspace{-2pt}
\subfloat{\includegraphics[scale=0.47]{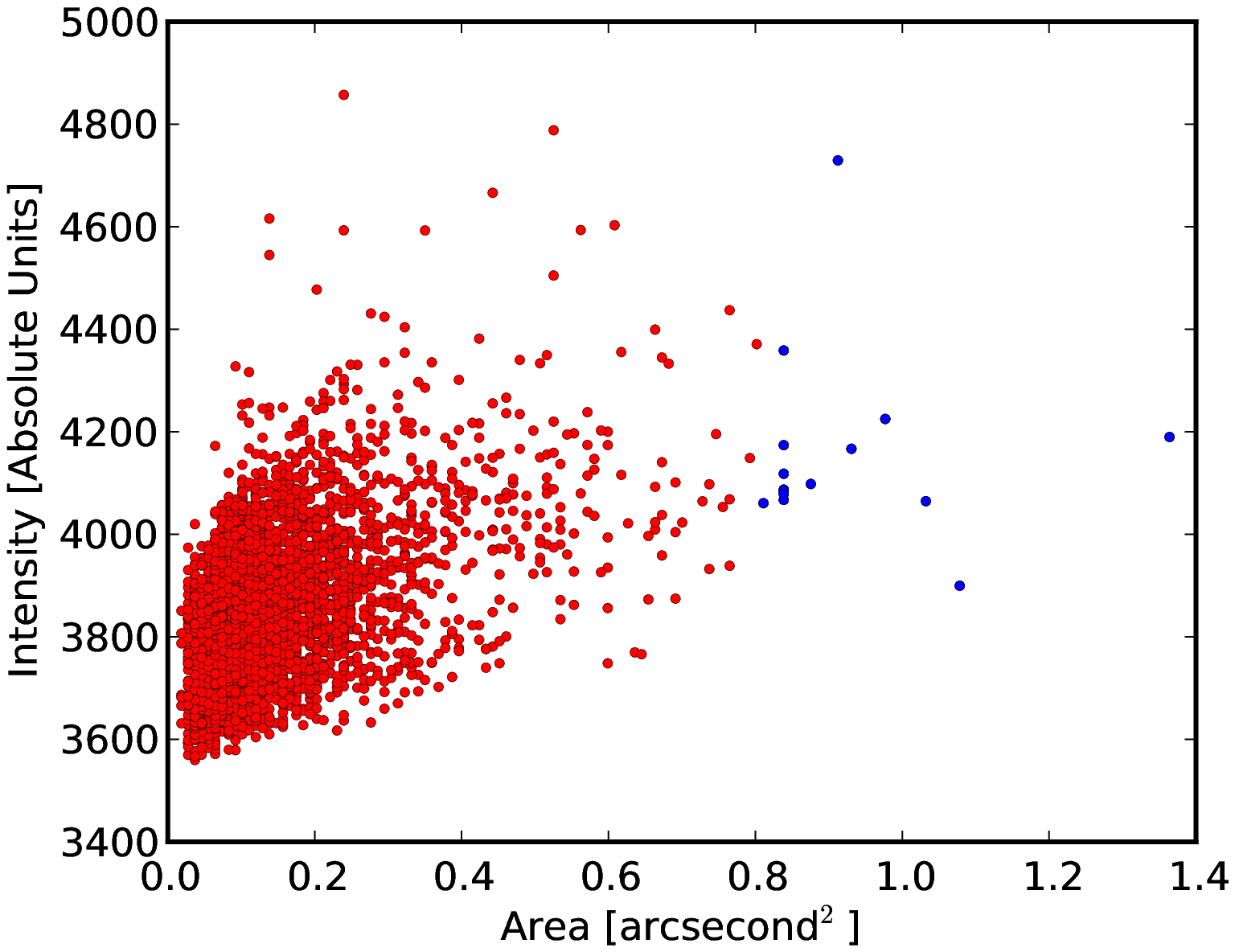}}
\vspace{-2pt}
\subfloat{\includegraphics[scale=0.5]{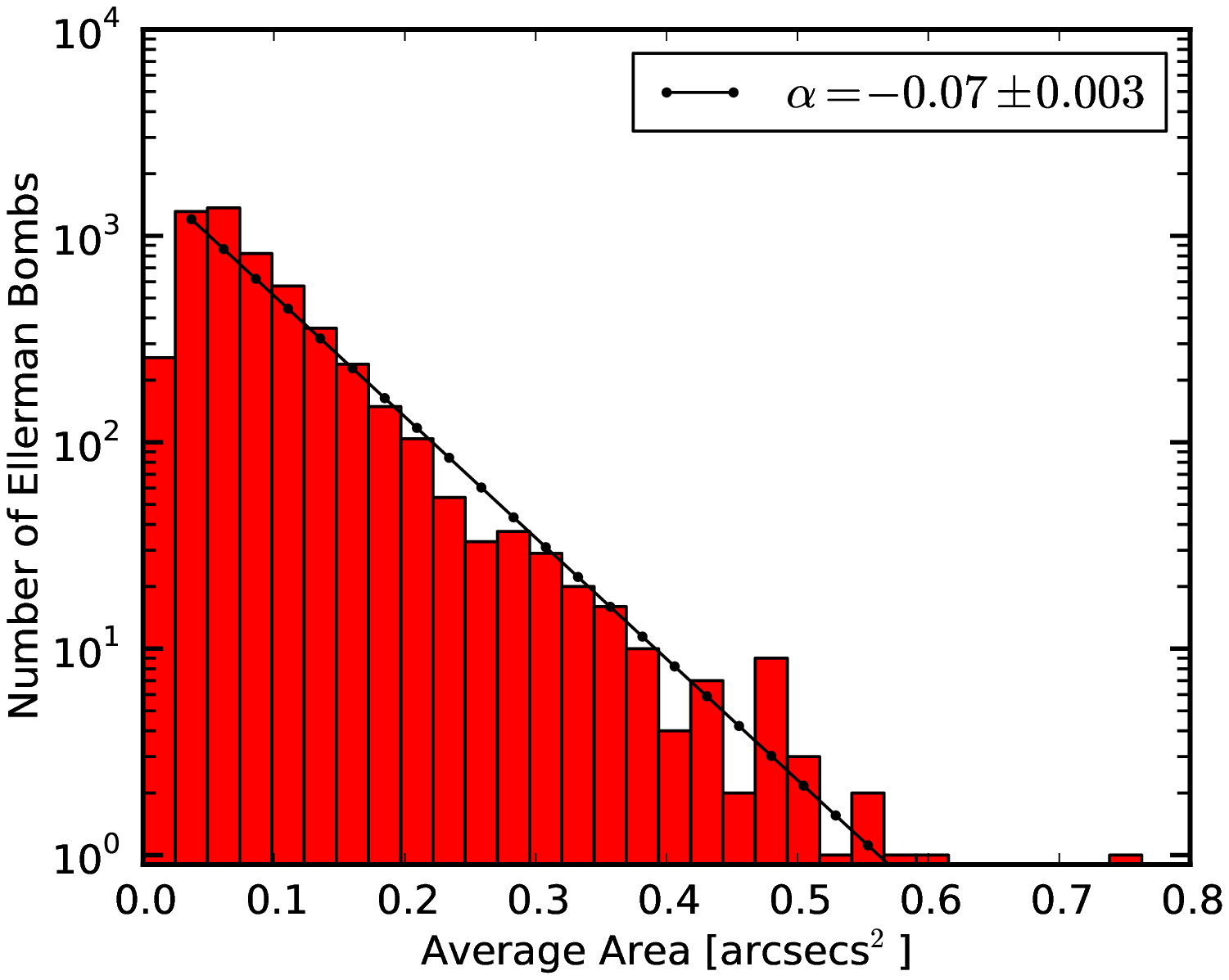}}
\caption{(a) Ellerman bomb lifetime frequency plot. { EBs with area larger than $0.8$\arcsec $\times 0.8$\arcsec\ (circular diameter of approximately $0.45$\arcsec) are shown using blue to compare estimates with previous researches}. (b) Intensity by area plot. Again, blue indicates larger events and red shows smaller events. (c) A histogram of area against log(frequency).}
\end{figure}

We suggest that EBs can be much shorter-lived than previous estimates and that in fact 
they can be { shorter} than our cadence. { By considering only the $3570$ EBs with lifetimes over one frame, we find an average lifetime of approximately three minutes which is considerably less than previous estimates of five to ten minutes. We find that for $5409$ EBs ({ including those EB events only observed in one frame}), the average lifetime drops to approximately two minutes suggesting that this research may only present an upper limit on lifetime.} { To check the reliability of the algorithm, large events (over the $0.8$\arcsec\ spatial-resolution presented by \opencite{Georgoulis02}) are also analysed (histogram plotted in blue in Figure 4(a)) giving an average lifetime for these events of approximately eight minutes}.   { A continued trend, as seen in Figure 4(a), coupled 
with Figure 4(b) could suggest that small magnetic reconnection events are extremely common in 
emerging active regions especially at the outer penumbra (see Figure 3) as well as the surrounding} { quiet Sun.} We suggest that a higher cadence dataset would find even shorter-lived EB events than presented here as well as better accounting for their recurrence.

\subsection{Area of Ellerman bombs}
EBs have been described as slightly elliptical brightenings with size of the order 
$1$\arcsec\ (for example \opencite{Zachariadis87}, \opencite{Georgoulis02}) such as in Figure 2. 
Our definition of EBs gives a large number of extremely small, less bright examples as 
shown in Figure 3. An intuitive explanation for this is that the spatial and temporal resolution that 
are used here, for example, are { one-fifth and one-seventh} of the Flare Genesis Experiment resolutions, 
respectively. We find that the majority of EBs in this study have diameters of below $0.5$\arcsec\ and find only a small number ({ $15$ over $0.8$\arcsec $\times {0.8}$\arcsec\ at some point in their lifetime}) with areas comparable to previous studies. 

The evolution of EBs over time is dominated by dynamic changes that were interpreted 
by \inlinecite{Watanabe11} as flaring. This flaring can consist of the rapid extension of thin ``arms'' 
away from the main structure or of a more general area expansion. We suggest that EBs that have an area 
of the order of our spatial resolution may still flare but, if this is at the same ratio as larger 
flaring EBs, this will not be picked up by these observations. { By analysing large events we do find some evidence of flaring within EBs through their lifetime; however, not all EBs, as identified in this study, can be seen to change visibly in either area or shape. We suggest that flaring may be a special case and not a defining feature of EB formation.}
 
Here, we find a plethora of small EB events (which are easily picked out in Figure 3 and are shown in 
Figures 4(b) and (c)) as well as many larger, more well-defined events. A positive correlation between the area 
and brightness of EBs is observed, implying the energy released from larger events is more powerful 
than the energy of smaller events. It should be noted that the small events have been missed due to 
both the spatial and temporal resolution of studies by {\it e.g.} \inlinecite{Georgoulis02} or \inlinecite{Pariat04}; however, with the spatial resolution in the current data, and the availability of the aligned 
SDO/HMI data (Figure 3), we are able to observe that many of these events are spatially linked to 
the approximate vertical magnetic field. This implies that many of the smaller events neglected in \inlinecite{Watanabe11} may be EB events on a smaller scale. In Section 3.6, we shall use high-resolution G-band data 
as a proxy for the magnetic field to discuss the formation of both the larger and smaller events.

\begin{figure}
\centering
\vspace{-20pt}
\subfloat{\includegraphics[scale=0.43]{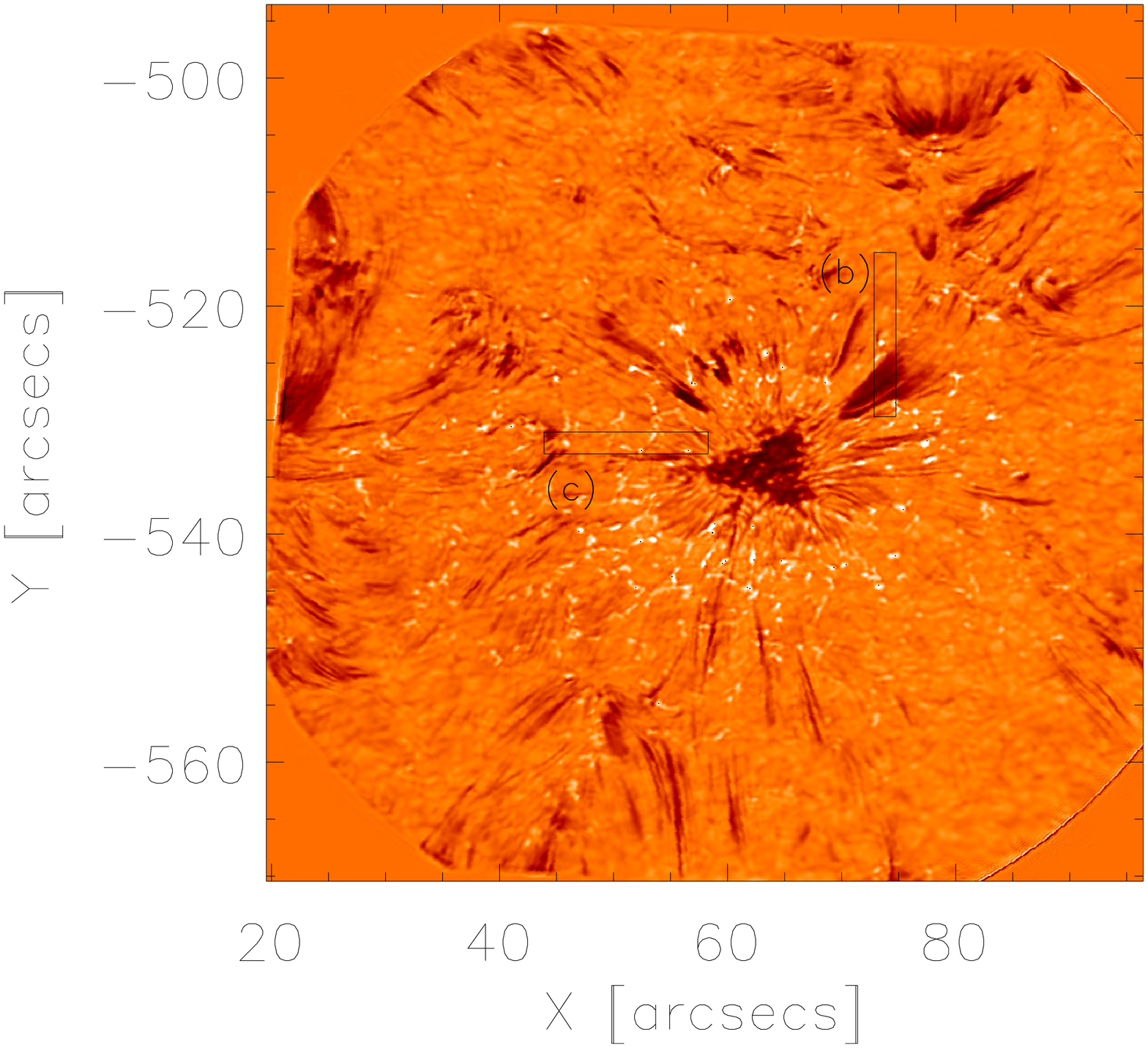}}
\vspace{-15pt}

\subfloat{\includegraphics[height=5 cm,width=12 cm]{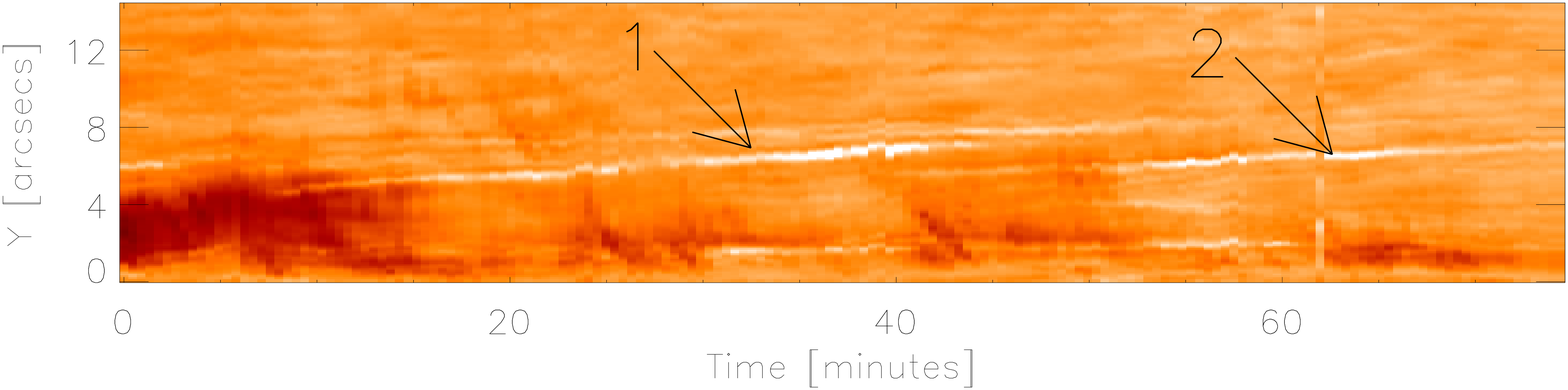}}
\vspace{-15pt}

\subfloat{\includegraphics[height=5 cm,width=12 cm]{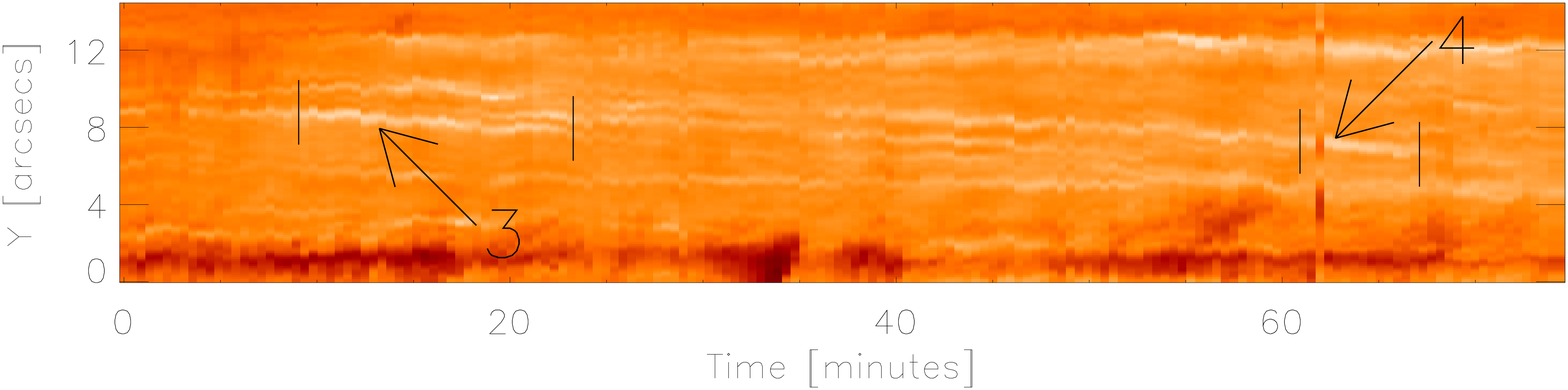}}

\caption{(a) Full FOV red wing image at time $=0$. (b) Space--time plot calculated by averaging the $x$-axis intensities over slit (b) ($20$ pixels wide) in the full FOV image. { Brightenings 1 and 2 occur from a similar spatial position and then migrate away from the spot}. (c) Space--time plot corresponding to slit (c) from the full FOV image.}
\end{figure}

\subsection{Movement and recurrence of Ellerman bombs}
The formation of Ellerman bomb events due to the movement of G-band magnetic 
bright points has been discussed by \inlinecite{Jess10}, who found that MBPs that migrate along the intergranular 
lanes can form the foot-points of H$\alpha$ micro-flare events. They reported a constant velocity of $1$ km s$^{-1}$ 
and $1.8$ km s$^{-1}$ for two G-band MBPs, which is matched well to the estimated MBP horizontal velocity stated by 
\inlinecite{Utz10}. \inlinecite{Keys11} extended this work, discussing the differences between isolated MBPs and 
merging MBPs. Interestingly, their results showed agreement with those found by \inlinecite{Utz10} in that isolated MBPs 
have an average velocity of around $1$ km s$^{-1}$; however, they found that merging MBPs appeared to have significantly 
larger velocities. { Should MBPs and EBs possibly be linked, as discussed previously, then} { one should expect motions to also be seen in EB events.} \inlinecite{Georgoulis02} and \inlinecite{Watanabe11} both studied EB movements within the 
H$\alpha$ line wings, finding horizontal motions of the order of 0.4 to 0.8 km s$^{-1}$. Here, we discuss the velocities of typical EB motions associated with recurring events (presented in Figure 5).

Figure 5(b) shows the horizontal motions of two EBs that occur from a similar spatial position and then migrate away from the sunspot with characteristic speeds around $1$ km s$^{-1}$. The brightenings form and become more pronounced over time becoming defined as EBs (over $130\ \%$ of the average image intensity) before fading away. We found that around half an hour after the initial brightening there is a second occurrence from the same position. Each trail is only characterised as an Ellerman bomb for a short amount of time due to rapid brightening and fading. We mark the first brightening with ``$1$'' and the second with ``$2$''. The recurrence of an EB from a similar spatial position could be interpreted as a continued emergence of flux leading to multiple reconnection and, therefore, EB events. 

{ The lower panel, Figure 5(c), shows the second definition of recurrence that is prevalent in these observations. We find many examples of extreme brightenings, classified as EBs, fading before becoming brighter again. Bars are situated around the EBs indicated by arrows $3$ and $4$ to show the extent of the lifetime of each event. One should note, however, that this is an x{\em -}t plot and, therefore, does not accurately convey relative brightness ({\it e.g.} if a full $20$ pixels are around $110\ \%$ of the background brightness, this may appear similar to a ten pixel $130\ \%$ brightness). To make this plot, we analysed a movie of the same period over time and noted that the EBs that are indicated by the bars appear to be connected. We find that many EBs tend to fade before regaining brightness at a later time, perhaps due to a recurrence of a triggering event such as magnetic reconnection. Over the course of its lifetime this EB/sustained brightening has a velocity of approximately $0.4$ km s$^{-1}$, which is consistent with velocities found by \inlinecite{Georgoulis02} and \inlinecite{Watanabe11}.}

\subsection{ Correlation Between G-band Magnetic Bright Points and EBs}
Current theories of EBs suggest that they are likely to be formed through magnetic reconnection in the lower solar atmosphere. In this Section we use G-band/ROSA data as a proxy for the magnetic field (\opencite{Berger01}; \opencite{deWijn09}) in the photosphere to further discuss this idea. MBPs observed in the G-band wavelength are created by the build-up of magnetic flux within inter-granular lanes deposited by their associated down-drafts, thereby leading to significant brightenings. A connection between 
G-band MBPs and EBs is discussed by \inlinecite{Jess10} who found 
that brightening events in the H$\alpha$ line wings can be spatially correlated to MBPs in the G-band. We 
expand on their work here by discussing examples of both large and small EBs and their relation to MBPs. 

We define a large EB to have an area of approximately $0.8$\arcsec $\times 0.8$\arcsec, the approximate average EB size in {\it e.g.} \inlinecite{Georgoulis02} and \inlinecite{Watanabe11}. By co-aligning the H$\alpha$ line-wing 
and G-band datasets, we find that there is a correlation between many EB brightenings and MBPs. EB $A$ in 
Figure 6, for example, is a large EB situated over an inter-granular lane and linked to photospheric bright points in the G-band image. { As has 
been shown by {\it e.g.} \inlinecite{Jess10}, large EBs can possibly be caused by interactions between two photospheric MBPs within inter-granular lanes.} { Note, as this sunspot is south of the disk-centre, MBPs will not appear within the inter-granular lanes; rather the southern walls of the event will appear below the dark lanes as with events $A$, $B$ and $C$.}

{ Examples exist within this FOV that do not correlate to brightenings within the G-band image (most noticeably at ($51$\arcsec, $-541$\arcsec). We suggest that this is due to MBPs being a possible proxy for strong vertical magnetic field within G-band data and EBs potentially being the product of magnetic reconnection, which does not neccesarily occur only at sites of vertical field. This implies, therefore, that G-band MBPs can lead to the formation of EBs, although, not in every case, and that EBs can form without interaction between MBPs. This supports models suggested by, for example, \inlinecite{Georgoulis02}, whereby multiple magnetic topologies can lead to EB formation.}

\begin{figure}
\centering
\subfloat{\includegraphics[scale=0.33]{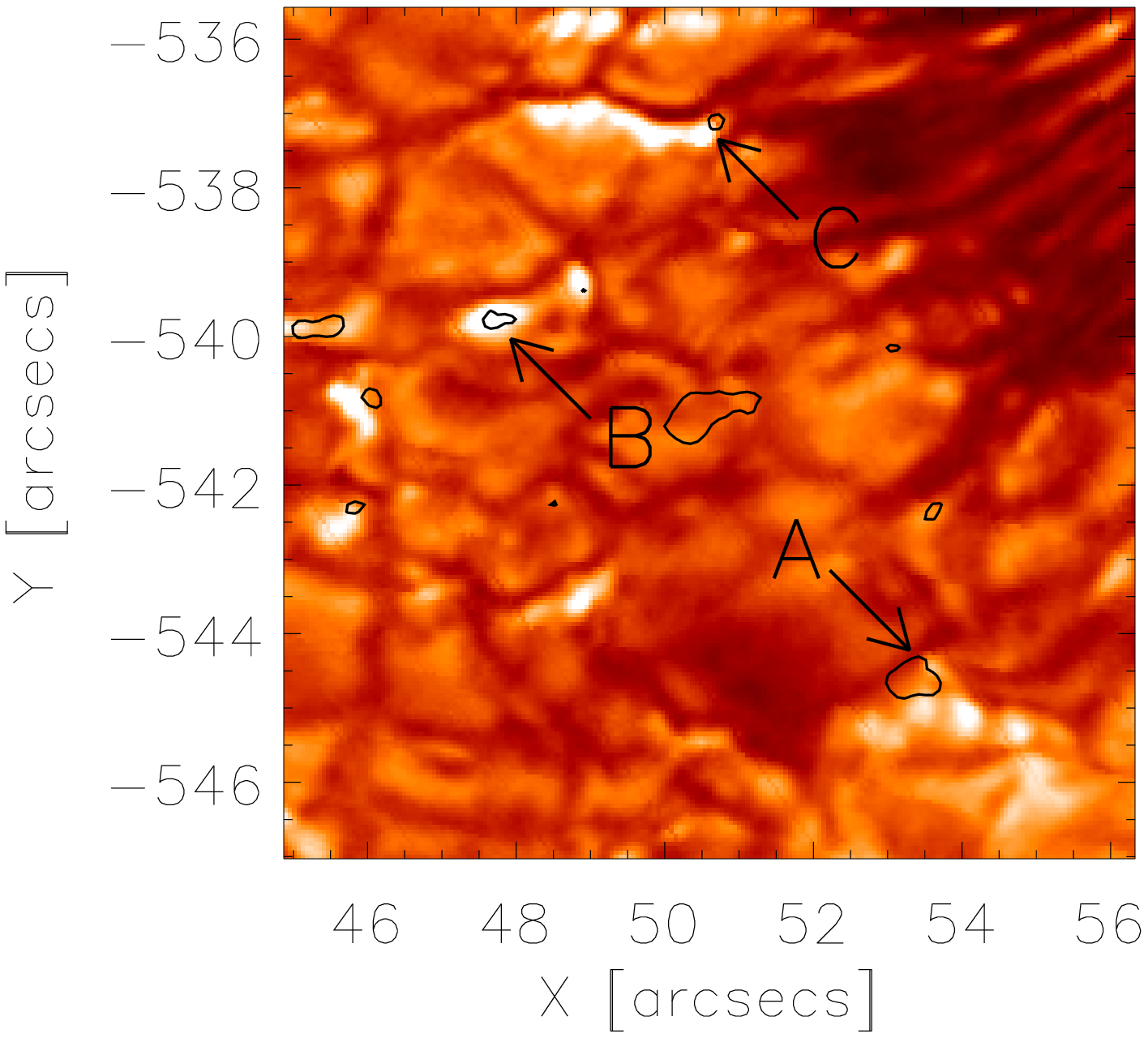}}
\subfloat{\includegraphics[scale=0.33]{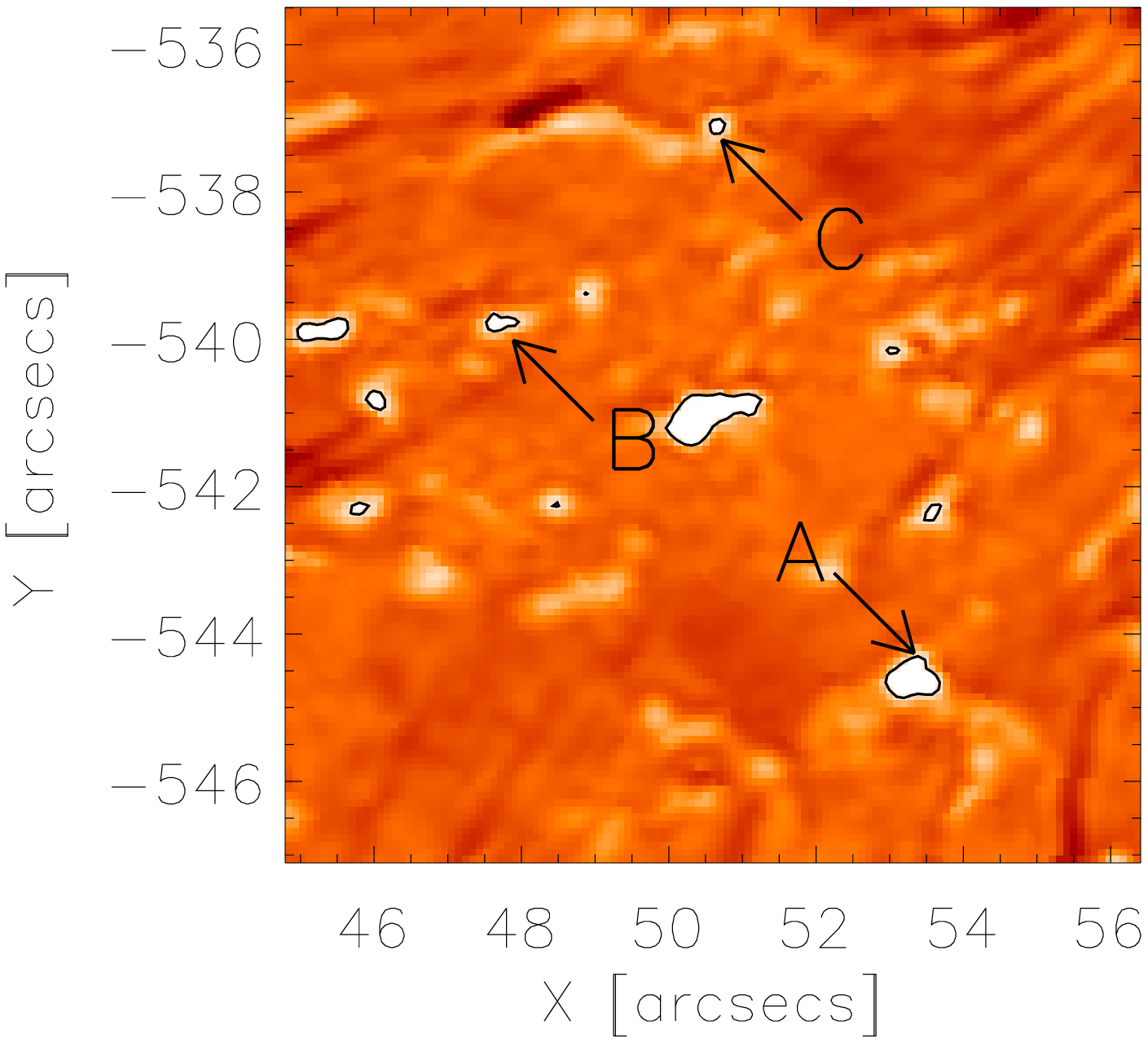}}
\caption{(a) G-band image with EBs contoured over it in white. Correlation between many brightenings is evident; three characteristic examples are shown with arrows. (b) Temporally aligned H$\alpha$ blue wing image with the arrows pointing to the EBs shown in (a).}
\end{figure}

We now discuss smaller EB events such as $B$ and $C$ from Figure 6.  The co-aligned data allow us to observe that these smaller EB events can also be linked to MBPs within the G-band dataset. EB $C$ does not increase in size from this image throughout its short lifetime. These two typical examples show a similar spatial relationship between MBPs within intergranular lanes and small EB events as that possessed by larger EB events. { It has been reported that two MBPs interacting can create EB brightenings in the H$\alpha$ line wings; however, for events $B$ and $C$, we see no evidence of fragmented MBPs ({\it i.e.} only one MBP is observable). Whether fragmenting occurs on scales smaller than the spatial-resolution presented here will be answered by higher-resolution data.}

We suggest that the potential link between small EBs and MBPs supports the assertion that the smaller events, which have been neglected in other studies, are indeed EB events. \inlinecite{deWijn09} proposed that magnetic structuring within the photosphere should happen on a spatial scale well below the diffraction limit of current telescopes; this has been supported by numerical simulations run by \inlinecite{Crockett10} who found the mode MBP size to be around $45\ 000$ km$^{2}$ (or a circular diameter of approximately $0.32$\arcsec) and an equal number of events with sizes $10\ 000$ km$^{2}$ and $100\ 000$ km$^{2}$ (circular diameters of $0.16$\arcsec\ and $0.49$\arcsec, respectively). Therefore, if there is a connection between magnetic structuring and EBs as widely anticipated (\opencite{Georgoulis02}; \opencite{Pariat04}; \opencite{Jess10}; \opencite{Watanabe11}), then high-resolution data, collected during periods of excellent atmospheric seeing, as presented here, should find EBs to be, on average, extremely small which is supported by our results from Figures 4(b) and (c). Both the cadence and resolution of modern telescopes need to be improved to conclusively establish whether these smaller events have the same properties as larger EBs in terms of morphology; however, we have shown that these small EB events do appear to have the same driver from the lower photosphere and that they are, therefore, worthy of further study.

\subsection{Energetics}
We have found that EBs are numerous and intense in the wings of the H$\alpha$ line profile around the active region sunspot that we have studied. \inlinecite{Georgoulis02} estimated the energies of each EB at its peak, $P_{\mathrm{rad}}$, and in total, $E_{\mathrm{rad}}$; we shall conduct a similar analysis on the EBs identified by the algorithm used this work as follows:
\begin{equation}
P_{\mathrm{rad}}  \approx \epsilon_{\mathrm{rad}}fV_{\mathrm{EB;max}} \quad{\textrm {and}}\quad E_{\mathrm{rad}} \approx \frac{P_{\mathrm{rad}}D}{2}
\end{equation}
where $\epsilon_{\mathrm{rad}}$ is the net radiative loss rate estimated by:
\begin{equation}
\epsilon_{\mathrm{rad}}  \approx {a(T)}n^2\chi{g(T)},
\end{equation}
$f$ is the radiative filling factor (assumed as being unity), $V_{\mathrm{EB;max}}$ is the maximum volume of the EB and $D$ is the lifetime of each EB. Defining $a(T)$ as the radiative reduction coefficient, $n$ as the total numerical density of electrons and neutral hydrogen, $\chi$ as the ionization degree and $g(T)$ as a semi-empirical function of the temperature from \inlinecite{Nagai80} and assuming $T \approx 10^4$ K and $n \approx 10^{12}$ cm$^{-3}$, \inlinecite{Georgoulis02} found $\epsilon_{\mathrm{rad}} \approx 0.89$ erg cm$^{-3}$ s$^{-1}$. Letting the height of an average EB be $10^2$ kilometres, we are able to produce estimates for the maximum and total energies of each of our EBs (Figure 7). 

\begin{figure}
\centering
\subfloat{\includegraphics[scale=0.65]{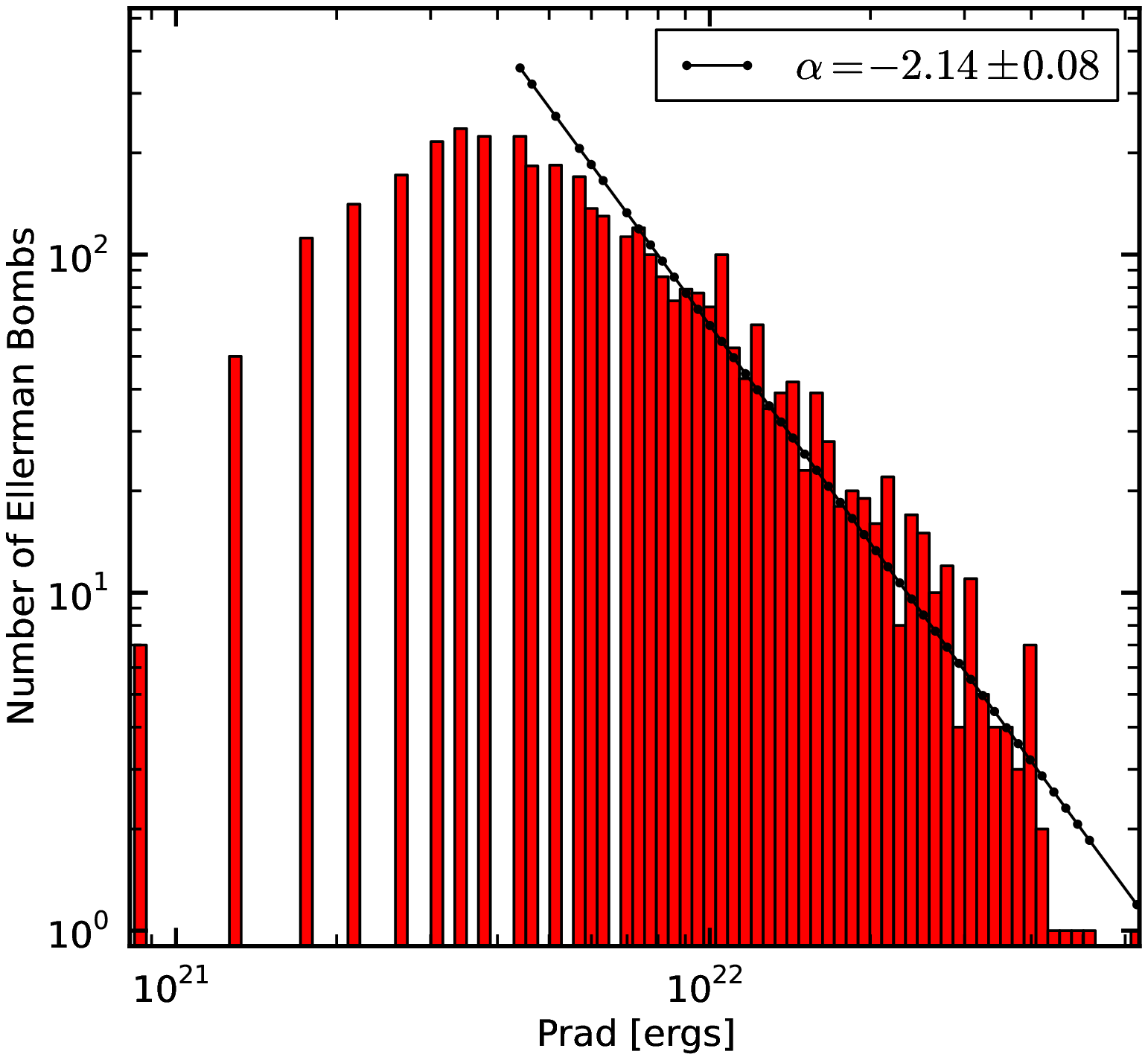}}
\vspace{-2pt}
\subfloat{\includegraphics[scale=0.65]{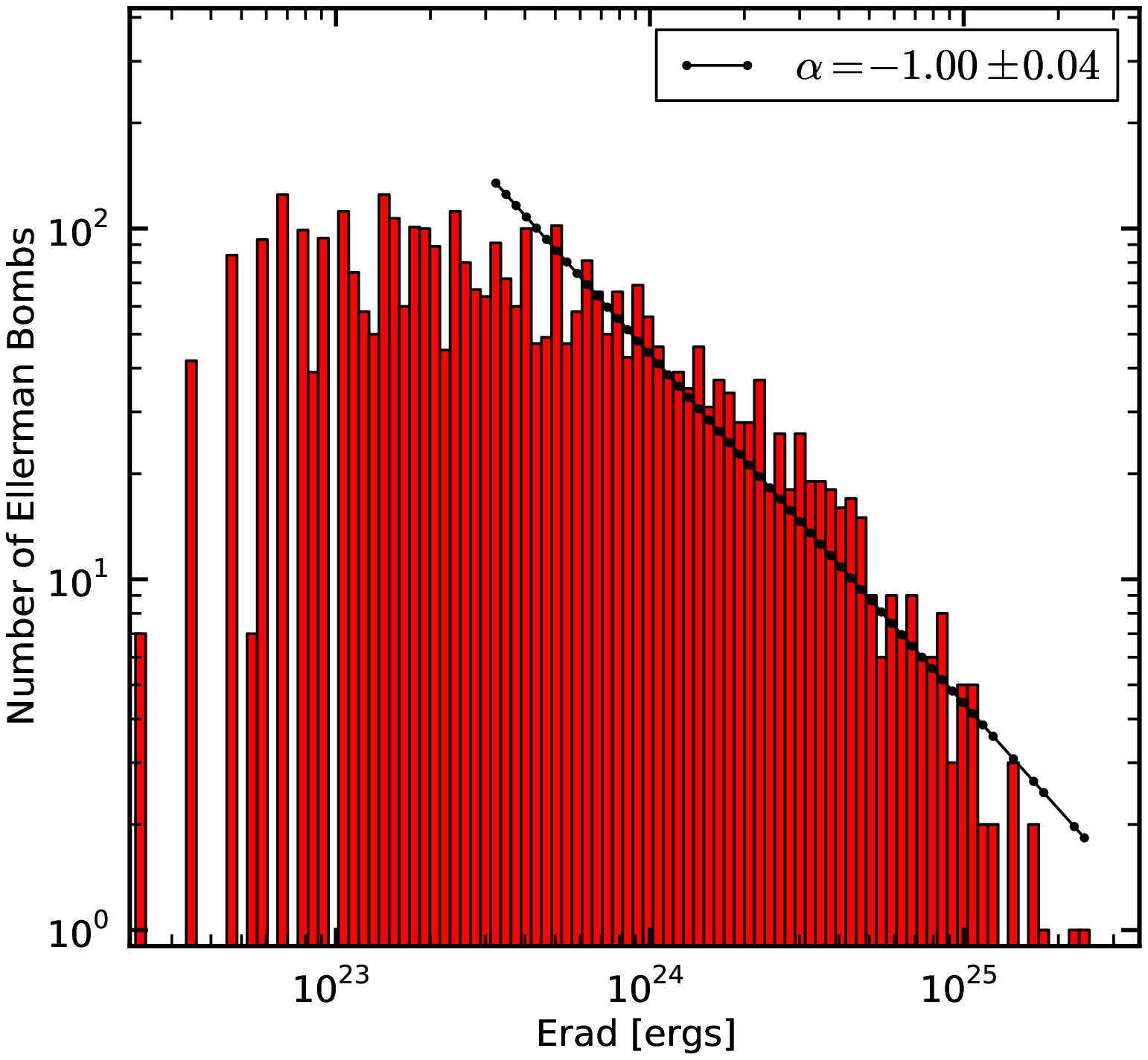}}
\caption{(a) The maximum energy of each EB. (b) The total energy release for each EB taking into account the estimated lifetime of each EB from our study.}
\end{figure}

We find that the energetics of the EBs studied in this article are around three to four orders of magnitude smaller than those stated in \inlinecite{Georgoulis02}. { A log-log plot of the number of EBs with respect to energy is plotted in Figures 7(a) and (b) appearing to show a power-law function ($\mathrm{d}N(x)/\mathrm{d}x\propto{x^{-\alpha}}$). Fits are plotted for both the total and maximum energy release graphs finding power law indices of  $\alpha_{\mathrm{total}}=-1.00\pm0.04$ and $\alpha_{\mathrm{max}}=-2.14\pm0.08$ respectively.} Interestingly, the total energy release histogram shows energies in the region $[2\times10^{22},\ 4\times10^{25}]$ ergs, which has been suggested as the possible energies of ``nano-flares'' by \inlinecite{Parnell00}. { We note that the power law that fits the total energy release does not imply that EBs are a major contributor to coronal heating; however, we suggest that this relation could be changed by higher temporal and spatial resolution data ({\it i.e.} by only including larger, longer-lived events where limitations in the observational resolutions are less evident, the power law changes so that $\alpha_{\mathrm{total}}>2$} { and, therefore, EBs have the required energy to sustain the corona).} 

{  We firmly remind the reader that these values are based on a large number of, possibly erroneous, assumptions such as the height over which EBs form, the temperature and total numerical densities stated here. For example, the temperature given in this work, $T \approx 10^4$ K, is, perhaps, a factor of two larger than average photospheric temperatures. The individual brightness of each event is also not included in this method, therefore not differentiating between extremely bright, high-energy events, and less bright, low-energy events. Overall, we suggest that the estimates of EB energies presented in this article should be a topic of consideration for future studies. The }

\section{Conclusions}
     \label{S-Conclusions}
The strong magnetic fields in the solar atmosphere are believed to be one of the key components that link the photosphere and the lower atmosphere to the corona and upper atmosphere. Here, we have discussed EBs, one possible manifestation of magnetic reconnection in the { lower-atmosphere}, concentrating on their morphology and statistical properties.

We find that the inclusion of ``flaring'' in the definition of EBs leads to the omission of many small-scale (less than $0.8$\arcsec) events that have the same signatures as larger, more widely studied EB brightenings in the H$\alpha$ line wings. Analysing the lifetimes and area of over $3500$ EB events, we find an average lifetime of $2-3$ minutes, much less than previous estimates, which have ranged from $5-15$ minutes ({\it e.g.} \opencite{Georgoulis02}; \opencite{Pariat04}; \opencite{Watanabe11}) and also a decreased average area. { To test the algorithm, events over the spatial resolution of previous researches were also analysed, allowing us to retrieve a lifetime estimate comparable to previous studies that did not benefit from the high-resolution, high-cadence data that are used here.}

Additionally, we find that the underlying G-band photospheric data show corresponding brightenings 
to intense H$\alpha$ brightenings even at a scale around $0.2$\arcsec. These smaller events also show { diminished} energies from previous studies with our dataset yielding energies of between $[2\times10^{22},\ 4\times10^{25}]$ ergs for the $3570$ EBs studied. { These energies also hint at a power law shape, which should be expanded upon by further studies.} This agrees well with the  theory presented by \inlinecite{deWijn09}, who suggested that magnetic reconnection and structuring  can happen well below current observational resolution and cadence. The continued advancement of  observational instruments should lead to more detailed studies of these small events to discuss their morphology in comparison to larger events. 

Overall, we find the EBs are short-lived, small-scale events that are common in the H$\alpha$ line wings around active region sunspots. They form over bipolar regions with complex mixing magnetic polarities  (as inferred from the SDO/HMI instrument) as well as within and around the edges of strong penumbrae (either through unipolar magnetic field interactions within the penumbra or bipolar fields in the surrounding region). { If evidence of magnetic reconnection leading to EBs in the lower-atmosphere is found, it is possible that they could prove an excellent driver of jets into the transition region and even, possibly, into the corona.}

\acknowledgements
Research at the Armagh Observatory is grant-aided by the N. Ireland Dept. of Culture, Arts and Leisure.  We 
thank the National Solar Observatory / Sacramento Peak for their hospitality and in particular Doug Gilliam for his
excellent help during the observations. We thank the UK Science and Technology Facilities Council for 
the studentships (CJN and SJM), PATT and support, plus support from grant ST/J001082/1. RE is thankful to the NSF, 
Hungary (OTKA, Ref. No. K83133) and acknowledges M. K\'eray for patient encouragement. We thank 
Friedrich W\"{o}ger for his Image Reconstruction code and S. B. Nicholson for the connected objects algorithm 
adapted for this study. The research leading to these results has received funding from the European Commission's 
Seventh Framework Programme (FP7/2007-2013) under the grant agreement eHeroes (project no. 284461, \url{http://www.eheroes.eu}). MM would like to thank the Air Force Office of Scientific Research, Air Force Material Command, USAF for sponsorship under grant number FA8655-09-13085. { HMI data courtesy SDO (NASA) and the HMI consortium.}

\bibliography{mybibstats}{}
\bibliographystyle{spr-mp-sola.bst}

\end{article}
\end{document}